\documentclass[12pt]{article}

\usepackage{amssymb}
\usepackage{latexsym}
\usepackage{cite}
\usepackage{graphicx}

\normalbaselines

\catcode `\@=11
\@addtoreset{equation}{section}

\def\section{\@startsection {section}{1}{\z@}{-3.5ex plus -1ex minus 
-.2ex}{2.3ex plus .2ex}{\normalsize\bf}} 
\def\subsection{\@startsection{subsection}{2}{\z@}{-3.25ex plus -1ex minus -.2ex}{1.5ex plus .2ex}{\normalsize\bf}}
\def\subsubsection{\@startsection{subsubsection}{3}{\z@}{-3.25ex plus -1ex minus -.2ex}{1.5ex plus .2ex}{\normalsize\sl}} 

\catcode`\@=12

\newcommand{\be}{\begin{equation}}
\newcommand{\en}{\end{equation}}

\newcommand{\bea}{\begin{eqnarray}}
\newcommand{\ena}{\end{eqnarray}}

\newcommand{\beano}{\begin{eqnarray*}}
\newcommand{\enano}{\end{eqnarray*}}

\newcommand{\bee}{\begin{enumerate}}
\newcommand{\ene}{\end{enumerate}}

\newcommand{\bei}{\begin{itemize}}
\newcommand{\eni}{\end{itemize}}

\def\C{{\mathbb C}}
\def\R{{\mathbb R}}
\def\N{{\mathbb N}}
\def\Z{{\mathbb Z}}

\newcommand{\nn}{\nonumber}
\newcommand{\noi}{\noindent}
\newcommand{\dis}{\displaystyle}

\newcommand{\da}{^\dagger}
\newcommand{\id}{1\raisebox{0.67mm}{\hspace*{-1mm}$\scriptstyle |$}}

\begin{document}

\thispagestyle{empty}

\vspace*{1cm}
\begin{center} 
{\large \textbf  {Temporally stable coherent states for infinite well }} \vspace{3mm}\\
{\large  \textbf { and P\"oschl-Teller potentials}} 
\vspace{1cm}
\\
 J-P. Antoine \footnotemark 
\\ 
{\small\it Institut de Physique Th\'eorique, Universit\'e Catholique de Louvain}\\
{\small\it  B - 1348  Louvain-la-Neuve, Belgium} \bigskip
\\
 J-P. Gazeau \footnotemark\  and P. Monceau  \footnotemark  
\\
{\small\it Laboratoire de Physique Th\'eorique de la Mati\`ere Condens\'ee}\\ 
{\small\it  Universit\'e Paris 7 -- Denis Diderot,
   F - 75251 Paris Cedex 05, France}\bigskip
\\
 J. R. Klauder  \footnotemark    
\\
{\small\it Departments of Physics and Mathematics,
  University of Florida, Gainesville, FL 32611, USA}\bigskip
\\
  K. A. Penson  \footnotemark  
\\
{\small\it Laboratoire de Physique Th\'eorique des Liquides,
 Universit\'e Paris 6 -- Pierre et Marie Curie}\\
{\small\it  F - 75252 Paris Cedex 05, France} 
\end{center}

\vspace{1cm}
\begin{abstract}
This paper is a direct illustration of a construction of coherent states which has been recently proposed by two
of us (JPG and JK). We have chosen the example of a particle trapped in an infinite square-well and also in
P\"oschl-Teller potentials of the trigonometric type. In the construction of the corresponding coherent states,
we take advantage of the simplicity of the solutions, which ultimately stems from
the fact they share a common $SU(1,1)$ symmetry {\it \`a la} Barut--Girardello. Many properties of these states are
 then studied,
both from mathematical and from physical points of view. 

\vfill

\bigskip\bigskip

\bigskip


\footnotetext[1]{Electronic address: {\tt antoine@fyma.ucl.ac.be}}
\footnotetext[2]{Electronic address: {\tt gazeau@ccr.jussieu.fr}}
\footnotetext[3]{Electronic address: {\tt pmo@ccr.jussieu.fr}}
\footnotetext[4]{Electronic address: {\tt klauder@phys.ufl.edu}}
\footnotetext[5]{Electronic address: {\tt penson@lptl.jussieu.fr}}

\end{abstract}

\section{INTRODUCTION}

Despite its relevance for the understanding of the most elementary parts of quantum mechanics, the
problem of a particle trapped in an infinite square-well (Figure \ref{figure1}) usually deserves no more than a few pages in
most physics textbooks \cite{atk,lali,flugge}.  Solutions are straightforward to derive, energy is nicely quantized
and trigonometric wave functions afford an immediate intuition of quantum behavior. The model is
widely used to give a fair idea of many body systems in atomic or molecular physics. However,
very soon one may become puzzled by less trivial problems pertaining to the mathematics of quantum
mechanics: domain of self-adjointness for the operators involved, possible nonuniqueness of
self-adjoint extensions (see in particular the very instructive examples in Chapters VIII.I, VIII.2,
and X.1 of  \cite{simon}), explicit kernel of the evolution operator, crucial role played by the
boundary conditions, semi-classical behavior and the classical limit, and other limiting situations such as
a very large or a vanishingly small width of the well.

Actually all these questions can be considered through a nice analytic regularization of the infinite well
potential. Indeed, consider the continuously indexed family of potentials 
\be
V(x) \equiv V_{\lambda, \kappa} (x) =
\frac{1}{2} V_o \left(\frac{\lambda (\lambda-1)}{\cos^2 \frac{x}{2a}} + 
\dis\frac{\kappa (\kappa -1)}{\sin^2\frac{x}{2a}} \right), \quad 0\leqslant x \leqslant \pi a,
\label{PTpot}
\en
for $\lambda, \kappa > 1$ ($ V_o > 0$ is a coupling constant). Clearly this is a smooth approximation, for
$\lambda, \kappa \to 1^+$, of the infinite square-well over the interval $ [0, \pi a]$. These potentials, called
the P\"{o}schl--Teller (PT) potentials \cite{flugge,poschl},  are shown in   Figure \ref{figure2} for the values 
$(\lambda,\kappa) = (4,4), (4,8), (4,16)$, respectively.
In order to make contact with standard quantum mechanics on the whole line, 
there are two possibilities. Either one requires that $V(x) = \infty$ outside the
interval $ [0, \pi a]$, or one periodizes the potential, with period $\pi a$, and one imposes periodic boundary 
conditions at the points $\{ n \pi a, \, n
\in \Z \}$. But, since the walls separating the successive cells are impenetrable, one may also simply ignore these
extensions and consider only the interval $ [0, \pi a]$, which we shall do in the present paper.

The P\"{o}schl--Teller potentials share with their infinite well limit the nice property of being analytically
integrable. The reason behind this can be understood within a group-theoretical context: The family of potentials
(\ref{PTpot}) possesses an underlying dynamical algebra,  namely $\mathfrak{su}(1,1)$ and the discrete series
representations of the latter. We recall that the discrete series UIR's of $\mathfrak{su}(1,1)$ are labelled by a
parameter $\eta$, which takes its values in $\{\frac 12, 1, \frac 32 ,2, \ldots \} $ for the discrete series {\em
stricto sensu}, and in $ [\frac 12, + \infty)$ for the extension to the universal covering of the group $SU(1,1)$.
The relation between the P\"{o}schl--Teller parameters and $\eta$ is given by
$$
2 \eta - 1 = \lambda + \kappa,
$$
and the limit case $\lambda, \kappa \rightarrow 1^+$ corresponds to $\eta = \frac 32$.

Other approaches in the past led to the group $SU(2)$ as the dynamical group for the P\"{o}schl--Teller potentials,
when $\lambda + \kappa$ is an integer \cite{inomata}.  We emphasize here the fact that the
$SU(1,1)$ approach seems more natural, for it extends easily and {\em naturally} to noninteger values of
$\lambda + \kappa$.

\begin{figure}

\setlength{\unitlength}{1cm}
\centering
\begin{picture}(0,6)
\put(-2,2){\begin{picture}(10,6)

\put(0,0){\vector(1,0){5}}
\put(1,0){\line(0,1){3}}
\put(4,0){\line(0,1){3}}

\put(1,-0.3){\makebox(0,0){$0$}}
\put(4,-0.3){\makebox(0,0){$\pi a$}}
\put(5,-0.3){\makebox(0,0){$x$}}
\put(0.5,3){\makebox(0,0){$\infty$}}
\put(4.5,3){\makebox(0,0){$\infty$}}

\end{picture}}
\end{picture}
\caption{The infinite square-well potential.} 
\label{figure1}
\end{figure}
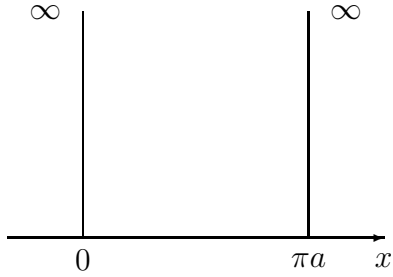
\begin{figure}
\begin{center}
\includegraphics[width=8cm]{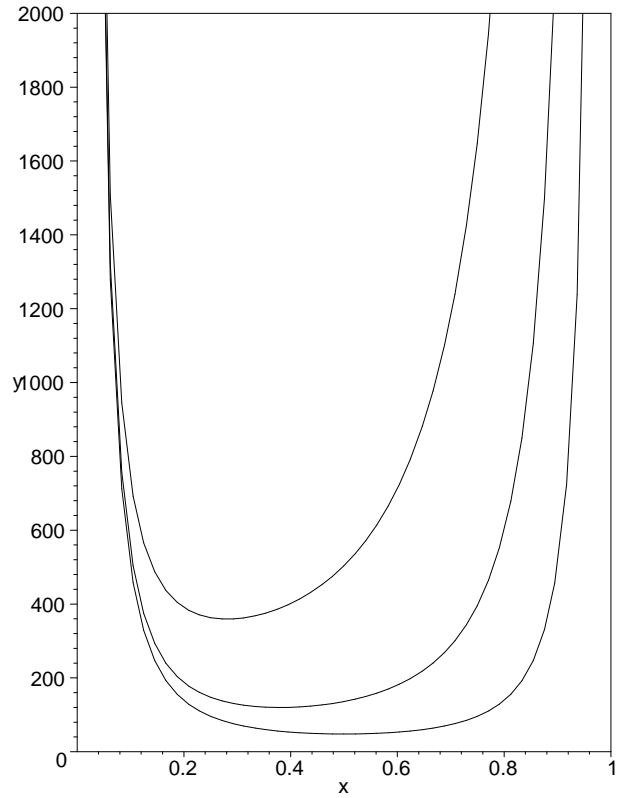} 
\end{center} 
\caption{The  P\"{o}schl--Teller potential 
$V(x) = \frac{1}{2} V_o \left[ {\lambda (\lambda-1)}{\cos^{-2} \frac{x}{2a}} + 
  {\kappa (\kappa -1)}{\sin^{-2}\frac{x}{2a}} \right],$ with $a =  \pi^{-1}$  
 and for $(\lambda,\kappa) = (4,4), (4,8), (4,16)$ (from  bottom to top).}
 \label{figure2}
\end{figure}

In fact, the P\"{o}schl--Teller potential (\ref{PTpot}), sometimes called PT of the first type, is closely
related to several other potentials, widely used in molecular and solid state physics.
\bei
\item
The symmetric P\"{o}schl--Teller potential well, given by $\lambda = \kappa \geqslant 1$,
\be
V_{\lambda}(x) = 2 V_o \dis\frac{\lambda(\lambda-1)}{\sin^2 \frac{x}{a}} \ . \label{symmPT}
\en
This potential may be periodized with period $\pi a/2$, instead of $\pi a$. 

\item
The same potential, for $ \frac 12 \leqslant \lambda < 1$, is known as the Scarf potential \cite{scarf}.  This is no
longer a well, but an inverted well, that is, a peak between two infinite negative wells. When periodized over the
whole line, this a good model for a 1-D crystal (as a smooth substitute to the well-known Kronig-Penney model), since
the spectrum of the corresponding Hamiltonian has a band structure. The nonsymmetric extension of the Scarf potential
has similar properties \cite{li-kus}.  
Interestingly, both cases admit $SU(1,1)$ as dynamical group, although the representations underlying the band
part are those of the complementary series.

\item
There exists also the so-called scattering (or modified) P\"{o}schl--Teller potentials, obtained by replacing the
trigonometric functions in (\ref{PTpot}) by their hyperbolic counter\-parts \cite{poschl}.  A special case is the
Rosen--Morse potential \cite{rosen-morse},  which is simply the symmetric version of the previous one. These
potentials are widely used in molecular physics, and they have the same dynamical group
$SU(1,1)$ (but again other representations are involved). For a review of this case and its applications, we refer to
  \cite{daskal1,alhass,frank_wolf}.
\eni 
In this paper we present and study families of coherent states (CS) adapted to the infinite well and to the
P\"{o}schl--Teller potentials. We call these states {\it adapted } and {\it coherent} because they are a direct
generalization of the standard ones corresponding to the harmonic oscillator \cite{klau1} (for an extensive and
up-to-date bibliography see, for instance,  
\cite{csbook}). We recall that the Schr\"{o}dinger--Klauder--Glauber CS read 
\be 
| z \rangle = e^{-\frac{\vert z
\vert^2}{2}}\sum_{n \geqslant 0} \frac {z^n}{\sqrt{n!}}\; | n\rangle.
\label{canCS}
\en
We extend them in a sense already explained in  \cite{klau2,ga-kl,ga-mon} and briefly sketched in the
following.
 We first consider in (\ref{canCS}) the kets $| n\rangle$ as the eigenstates of the infinite well (resp. P\"{o}schl--Teller)
Hamiltonian $H$ corresponding to the eigenvalue $ \hbar \omega e_n$, $n\geq 0, e_0 = 0$. Next, analogous to the
pioneering work of Jackson \cite{jackson},  we replace in the square root the factorial $n!$ by the generalized
factorial $[ e_n ] ! = e_1
\cdots e_n$, to get the so-called action identity \cite{ga-kl} 
\be
\langle z | H | z \rangle = \hbar \omega | z |^2. 
\en 
Note that similar factorial ``deformations'' in the construction of
coherent states already appear in   \cite{daskal2,sol}. We finally require (temporal) stability for our new
family of coherent states under the action of the evolution operator $e^{-iHt/\hbar}$ (see  \cite{ga-kl} and
Section VII below for details). Our interest for these infinite well and P\"{o}schl--Teller coherent states lies
mostly in the simplicity of the formulas involved. We have here at our disposal a nice tool for examining many
quantum features, such as probability densities, autocorrelation, mean values of observables, Heisenberg inequalites,
semi-classical limits, and others. 

The paper is organized as follows. In Section II, we describe the classical motion in an infinite square-well
potential and in a P\"{o}schl--Teller potential. In our opinion, it is essential to recall this elementary (and
pedagogical!) material for the subsequent discussions on a quantum level. Sections III and IV are devoted to the
quantum infinite well and P\"{o}schl--Teller potentials, respectively. In particular, we give here an up-to-date
survey of the nontrivial questions of self-adjointness for some of the most familiar physical observables. We examine
in Section V the questions related to various limits: semiclassical
$\hbar\, \rightarrow \, 0,\ n\hbar = \mbox{const}$, infinite narrowness $a \, \rightarrow \, 0$, infinite width $a \,
\rightarrow
\, \infty$, P\"{o}schl--Teller $\rightarrow$ infinite well, and others. We describe in Section VI the dynamical
symmetry algebra ${\mathfrak{su}}(1,1)$ common to both models and underlying their integrability. In Section VII we
first review the general construction of ``action--angle" or rather ``energy--time" coherent states before giving
their explicit form and their most immediate mathematical properties in the infinite well case, and in the
P\"{o}schl--Teller case (Section VIII).  Section IX is devoted to the most interesting physical properties of our
states, and in particular to the revival features they present, which are well illustrated by the large number of
figures shown there. Finally, Section X summarizes the discussion about the role of coherent states when expressed in
terms of action--angle variables.

A final lesson of the paper is that a comprehensive study of quantum mechanics requires not only algebra, or numerical
simulations, but also a precise use of functional analysis. The fine points of the latter are not mathematical
pedantry, they express deep physical properties.

\section{THE CLASSICAL PROBLEM}

\subsection{Classical infinite well}

It is worthwhile to start out this paper by a short pedagogical review of the classical behavior
of a particle of mass $m$ trapped in an infinite well of width $\pi a$.  

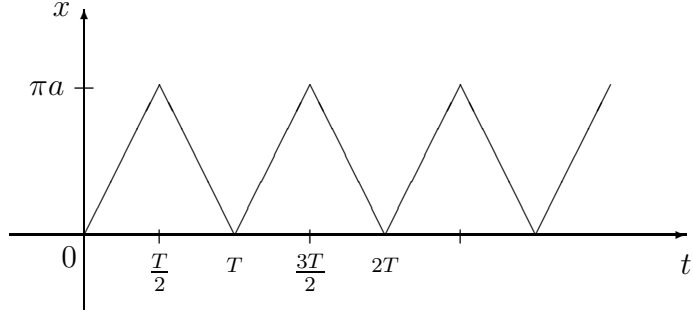
\begin{figure}
\centering
\setlength{\unitlength}{1cm}

\begin{picture}(0,6)

\put(-3.4,1){\begin{picture}(10,6)

\put(-1,0){\vector(1,0){9}}
\put(0,-1){\vector(0,1){4}}
\put(0,0){\line(1,2){1}}
\put(2,0){\line(-1,2){1}}
\put(2,0){\line(1,2){1}}
\put(4,0){\line(-1,2){1}}
\put(4,0){\line(1,2){1}}
\put(6,0){\line(-1,2){1}}
\put(6,0){\line(1,2){1}}

\put(-0.2,-0.3){\makebox(0,0){$0$}}
\put(-0.5,1.95){\makebox(0,0){$\pi a$}}
\put(0,1.95){\makebox(0,0){$-$}}
\put(-0.3,3){\makebox(0,0){$x$}}
\put(8,-0.4){\makebox(0,0){$t$}} 

\put(1,0){\makebox(0,0){$\shortmid$}}
\put(1,-0.5){\makebox(0,0){$\frac{T}{2}$}} 
\put(2,-0.4){\makebox(0,0){$\scriptstyle T$}} 
\put(3,0){\makebox(0,0){$\shortmid$}}
\put(3,-0.5){\makebox(0,0){$\frac{3T}{2}$}} 
\put(4,-0.4){\makebox(0,0){$\scriptstyle 2T$}} 
\put(5,0){\makebox(0,0){$\shortmid$}}

\end{picture}}
\end{picture}
\caption{The position $x(t)$ of the particle trapped in the infinite square-well of width $\pi a$, 
as a function of time.} 
\label{figure3}
\end{figure}

\begin{figure}
\centering

\setlength{\unitlength}{1.1cm}

\begin{picture}(0,6)

\put(-3,2){\begin{picture}(10,6)

\put(-1,0){\vector(1,0){8}}
\put(0,-1.5){\vector(0,1){4}}

\put(0,1){\line(1,0){1}}
\put(1,-1){\line(0,1){2}}
\put(1,-1){\line(1,0){1}}
\put(2,-1){\line(0,1){2}}
\put(2,1){\line(1,0){1}}
\put(3,-1){\line(0,1){2}}
\put(3,-1){\line(1,0){1}}
\put(4,-1){\line(0,1){2}}
\put(4,1){\line(1,0){1}}

\put(-0.2,-0.3){\makebox(0,0){$0$}}
\put(-0.4,2.5){\makebox(0,0){$\mbox{\boldmath $v$} (t)$}} 
\put(7,-0.4){\makebox(0,0){$t$}}

\put(0.8,-0.4){\makebox(0,0){$\frac{T}{2}$}} 
\put(1.8,-0.3){\makebox(0,0){$\scriptstyle T$}} 
\put(2.75,-0.4){\makebox(0,0){$\frac{3T}{2}$}} 
\put(3.8,-0.3){\makebox(0,0){$\scriptstyle T$}} 

\end{picture}}
\end{picture}
\caption{The velocity $\mbox{\boldmath $v$} (t)$ of the particle in the infinite square-well:
periodized Haar function.}
\label{figure4}
\end{figure}
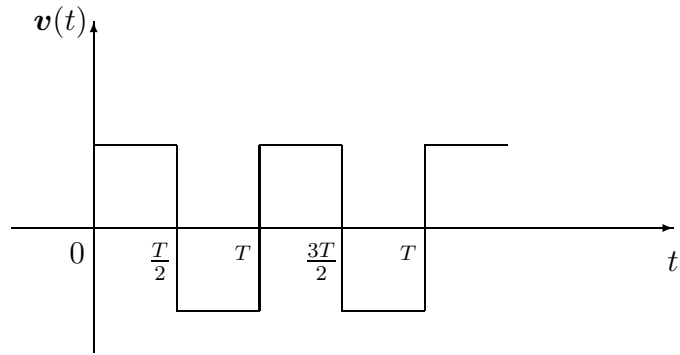

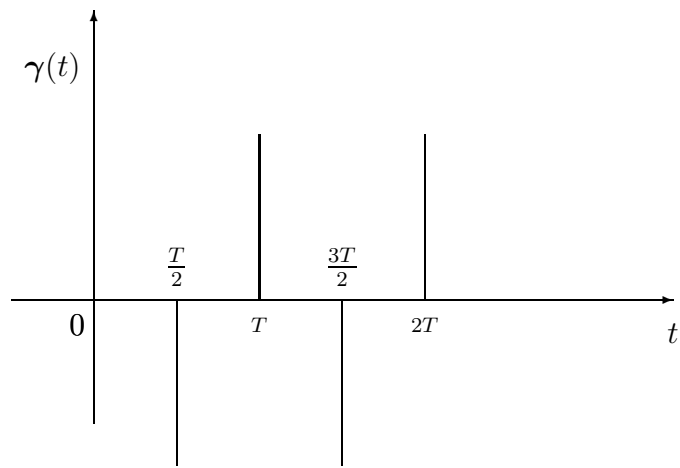
\begin{figure}
\centering

\setlength{\unitlength}{1.1cm}

\begin{picture}(0,6)

\put(-3,2){\begin{picture}(10,6)

\put(-1,0){\vector(1,0){8}}
\put(0,-1.5){\vector(0,1){5}}

\put(1,0){\line(0,-1){2}}
\put(2,0){\line(0,1){2}}
\put(3,0){\line(0,-1){2}}
\put(4,0){\line(0,1){2}}

\put(-0.2,-0.3){\makebox(0,0){$0$}}
\put(-0.2,-0.3){\makebox(0,0){$0$}}
\put(-0.5,2.8){\makebox(0,0){$\mbox{\boldmath $\gamma$} (t)$}} 
\put(7,-0.4){\makebox(0,0){$t$}}

\put(1,0.4){\makebox(0,0){$\frac{T}{2}$}} 
\put(2,-0.3){\makebox(0,0){$\scriptstyle T$}} 
\put(3,0.4){\makebox(0,0){$\frac{3T}{2}$}}
\put(4,-0.3){\makebox(0,0){$\scriptstyle 2T$}} 

\end{picture}}
\end{picture}
\caption{The acceleration \mbox{\boldmath $\gamma$(t)} of the particle
of the particle in the infinite square-well.} 
\label{figure5}
\end{figure}

For a nonzero energy
\be
E=\frac 12 mv^2,
\en
there corresponds a speed
\be
v =\sqrt{\frac{2E}{m}}
\en
for a position $0 < x <\pi a$.
There are perfect reflections at the boundaries of the well. So the motion is periodic
with period (the ``round trip time'') $T$ equal to 
\be 
T=\frac{2\pi a}{v} = 2\pi a \sqrt{\frac{m}{2E}}. 
\label{wellperiod}
\en
With the initial condition $x(0)=0$, the time behavior of the position is then given by (see Figure \ref{figure3}) 
\bea
0 \leqslant t \leqslant \frac 12 T &:\quad &x=vt, \nn\\ 
\frac 12 T \leqslant t \leqslant T &:\quad & x=2\pi a -vt, 
\label{wellevol}
\ena
and of course
\be
x(t+nT) = x(t).
\en
Consequently the velocity is a periodized Haar function (Figure \ref{figure4}): 
\be
\mbox{\boldmath $v$}   = v \sum_{n=0}^{+\infty}   
\left [{\id}_{[nT,(n+\frac12)T]}-{{\id}}_{[(n+\frac12)T,(n+1)T]}\right]  
\en
(here ${\id}_B$ denotes the characteristic function of a set $B\in \R$),
whereas the acceleration is the superposition of two Dirac combs on the half-line (Figure~\ref{figure5}):
\be
\mbox{\boldmath $\gamma$} = \sum_{n=0}^{+\infty} \left[\delta_{nT}-\delta_{(n+\frac12)T}\right]. 
\en

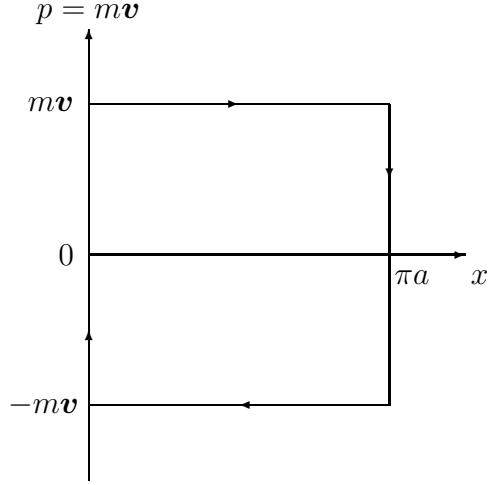
\begin{figure}

\centering

\setlength{\unitlength}{1cm}
\begin{picture}(16,8)
\put(6,3){\begin{picture}(10,6)

\put(0,0){\vector(1,0){5}}
\put(0,0){\vector(0,1){3}}
\put(0,2){\vector(1,0){2}}
\put(2,2){\line(1,0){2}}
\put(4,2){\vector(0,-1){1}}
\put(4,1){\line(0,-1){3}}
\put(4,-2){\vector(-1,0){2}}
\put(2,-2){\line(-1,0){2}}
\put(0,-3){\vector(0,1){2}}
\put(0,-1){\line(0,1){1}}

\put(-0.3,0){\makebox(0,0){$0$}}
\put(-0.5,2){\makebox(0,0){$m \mbox{\boldmath $v$}$}} 
\put(-0.6,-2){\makebox(0,0){$-m\mbox{\boldmath $v$}$}}
\put(0,3.2){\makebox(0,0){$p=m \mbox{\boldmath $v$}$}} 

\put(4.3,-0.3){\makebox(0,0){$\pi a$}}
\put(5.2,-0.3){\makebox(0,0){$x$}}

\end{picture}}
\end{picture}
\caption{Phase trajectory of the particle in the infinite square-well.} 
\label{figure6}
\end{figure}

\noindent
The average position and average velocity of the particule are then 
\be
\overline {x} = \frac{1}{T} \int_0^T x(t) \, dt = \frac{\pi a}{2}, \qquad \overline v = 0,
\label{av_pos}
\en
whereas the mean square dispersions are
\be
\sqrt{\overline {x^2}-\overline {x} ^2} = \frac{\pi a}{2\sqrt{3}} , \qquad 
\sqrt{\overline {v^2}-\overline {v} ^2} = \sqrt{\frac{2E}{m}} . 
\label{disper}
\en
Note the standard Fourier expansion for
the position and the velocity, respectively,
 \bea
 x(t) &=& \frac{\pi a}{2} - \frac{4a}{\pi} \sum_{n=0}^\infty \frac{1}{(2n+1)^2} \cos \frac{2\pi}{T}(2n+1)t, \\ 
\mbox{\boldmath $v$} (t) &=& 4\frac{v}{\pi}\sum_{n=0}^\infty \frac{1}{2n+1} \sin \frac{2\pi}{T}(2n+1)t.
\ena

Figure \ref{figure6} shows the phase trajectory of the system. This trajectory encircles a surface of
area equal to the action variable
\be
A = \frac{1}{2 \pi} \oint p dq = mva,
\en
where $q=x$ and $p=mv$ are canonically conjugate. Note the other expressions for $A$:
\be
A = \frac{2 \pi a^2 m}{T} = \frac{m v^2 T}{2 \pi} = a \sqrt{2mE}. \label{wellaction}
\en

The action-angle variables $(A,\varphi)$ are obtained through the canonical transformation
($\vartheta$ denotes the step function)
\bea
\varphi &=& {\rm sgn}(p) \frac{q}{a} +\vartheta (-p) 2 \pi \quad ({\rm mod}\, 2\pi),
\\
A &=& |p| a,
\ena
with generating function equal to the Maupertuis action, as should be expected: \bea
S_o &=& 2 \pi n A + S_o^{princ} \nn\\
S_o^{princ} &=& A \varphi, \quad \varphi \in (0,2 \pi) \nn\\ &=& pq + |p| \vartheta(-p)a.
\ena
Finally, note the time evolution of the angle variable: 
\be 
\varphi = \frac{\mbox{\boldmath $v$}}{a}t + \varphi_o \equiv \alpha t + \varphi_o. 
\en 

\subsection{P\"{o}schl--Teller potentials}

The solution to the equations of motion with the potentials (\ref{PTpot}) is straightforward,
in spite of the rather heavy expression of the latter. The turning points $x_\pm$ of the periodic motion
at a given energy $E$ are given by
\be
x_\pm = a\; \arccos \left[\frac{\alpha-\beta}{2} \pm \sqrt \Delta \right] \en where 
$\Delta = (1 - \frac 12 (\sqrt \alpha + \sqrt \beta)^2) (1- \frac 12 (\sqrt \alpha - \sqrt \beta)^2)$,
$\alpha = \dis\frac{V_0}{E} \lambda (\lambda -1)$, $\beta = \dis\frac{V_0}{E} \kappa (\kappa-1)$.
\medskip

So the motion is possible only if
\be
E > \frac{V_0}{2} (\sqrt{\lambda (\lambda -1)} + \sqrt{\kappa (\kappa -1)})^2. \label{motioncond}
\en
The time evolution of the position is given by \bea
x(t) &=& a\; \arccos \left[\frac{\alpha - \beta}{2} + \sqrt \Delta \cos(\sqrt{\frac{2E}{m}}\, \frac ta )\right]\ , 
\label{PTevol}
\\
x(0) &=& x_- \ . \nn
\ena
Hence the period is
\be
T = 2 \pi a\,\sqrt{\frac{m}{2E}}\ .
\label{PTperiod}
\en
It is remarkable that the period $T$ does not depend on the strength $V_o$, nor on $\lambda$ and $\kappa$.
\smallskip

The action variable $A$ satisfies the relation $\dis\frac{dA}{dE} = \dis\frac {T}{2\pi}$, and thus
\be
A = a \sqrt{2mE} + \mbox{ const }.
\en
The constant is determined by the condition that $A = 0$ for $E = V_{min}$, that is,
const = $- a \sqrt{2mV_{min}}$. The P\"{o}schl--Teller potential $V(x)$ reaches its minimum at the location $x_o$ defined by 
\be
\tan^2 \frac{x_o}{2a} = \sqrt{\frac{\kappa (\kappa -1)}{\lambda (\lambda-1)}}. \en
So we have, in agreement with (\ref{motioncond}), \be V_{min} = V(x_o) = \frac{V_o}{2}
\left[\sqrt{\lambda (\lambda-1)} + \sqrt{\kappa (\kappa -1)}\right]^2, \en
and consequently,
\be
A = a \sqrt{2mE} - a \sqrt{mV_o} [\sqrt{\lambda (\lambda-1)} + \sqrt{\kappa (\kappa -1)}].
\label{PTaction}
\en

\begin{figure} 
\begin{center}
\includegraphics[width=6cm,height=6cm]{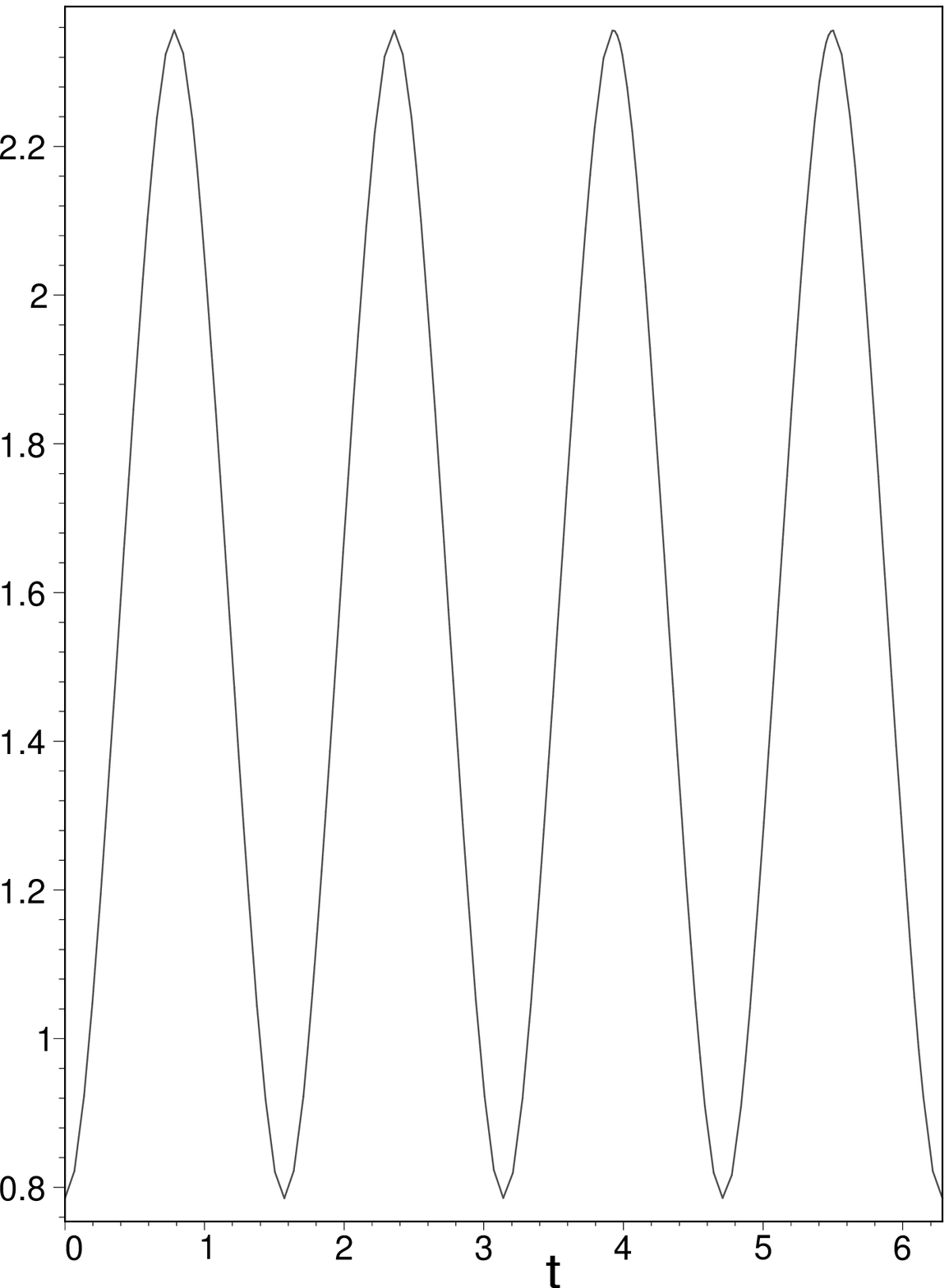}
\hspace{3mm}
\includegraphics[width=6cm,height=6cm]{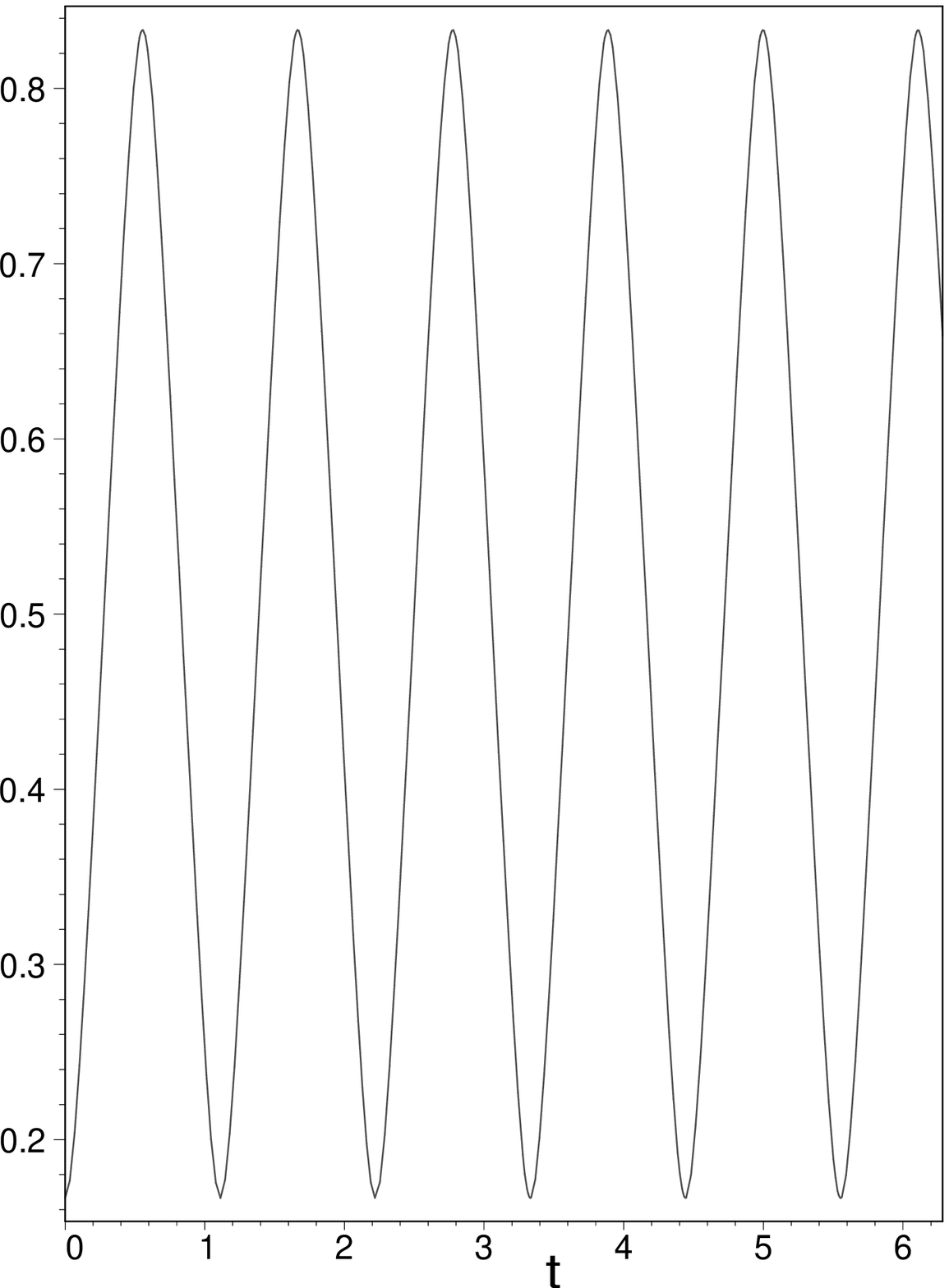}
\end{center}
\caption{The position $x(t)$ of the particle in the symmetric P\"{o}schl--Teller potential
$\lambda =\kappa = 2$: (a) $E=8 V_o, \,T = \frac{\pi}{2}$;  and (b) $E=16 V_o, \,T = \frac{\pi}{2\sqrt{2}}$
(compare   Figure \ref{figure3}).}
\label{figure7}
\end{figure}

\begin{figure}
\begin{center}
\includegraphics[width=6cm,height=6cm]{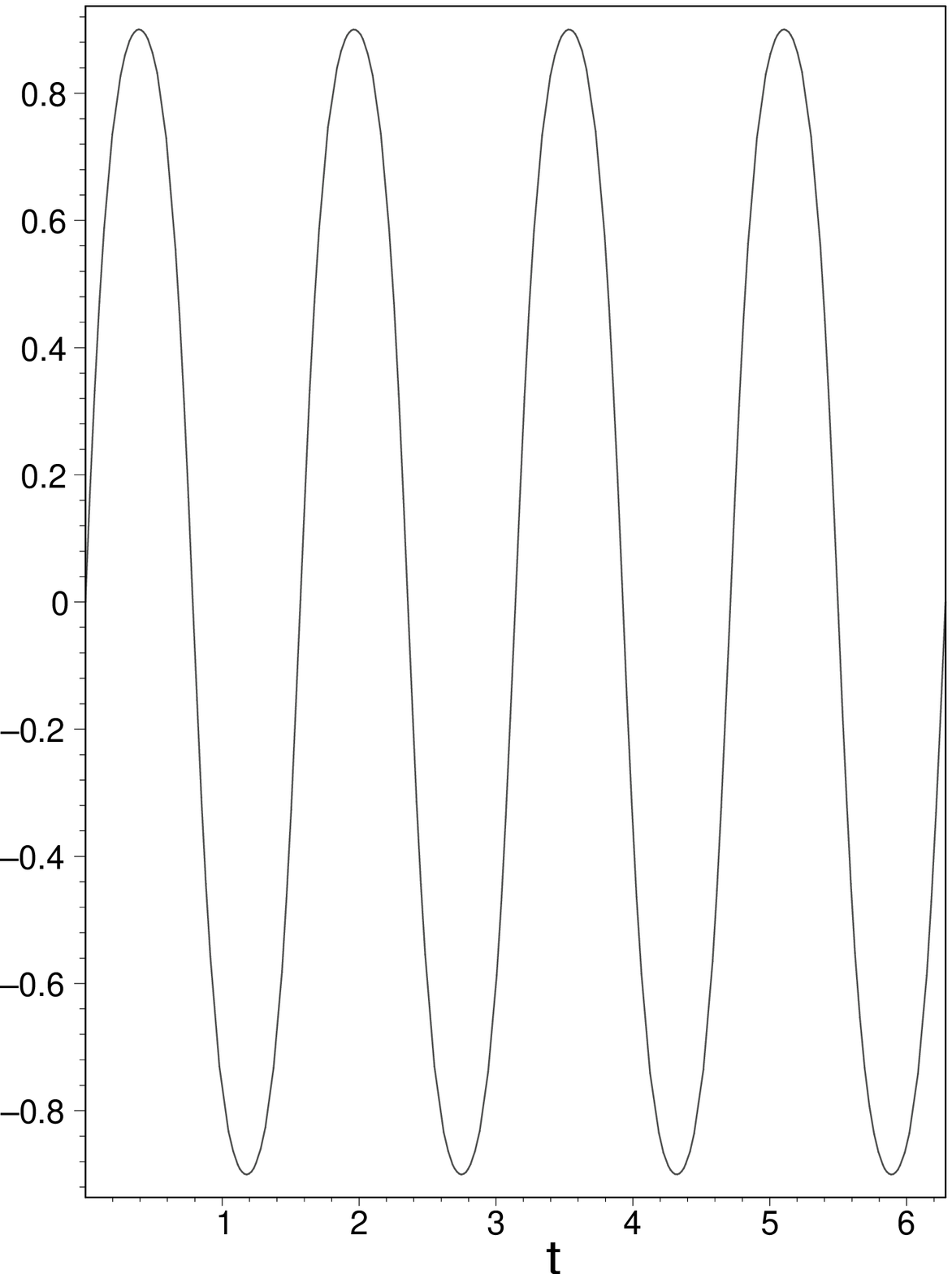}
\hspace{3mm}
\includegraphics[width=6cm,height=6cm]{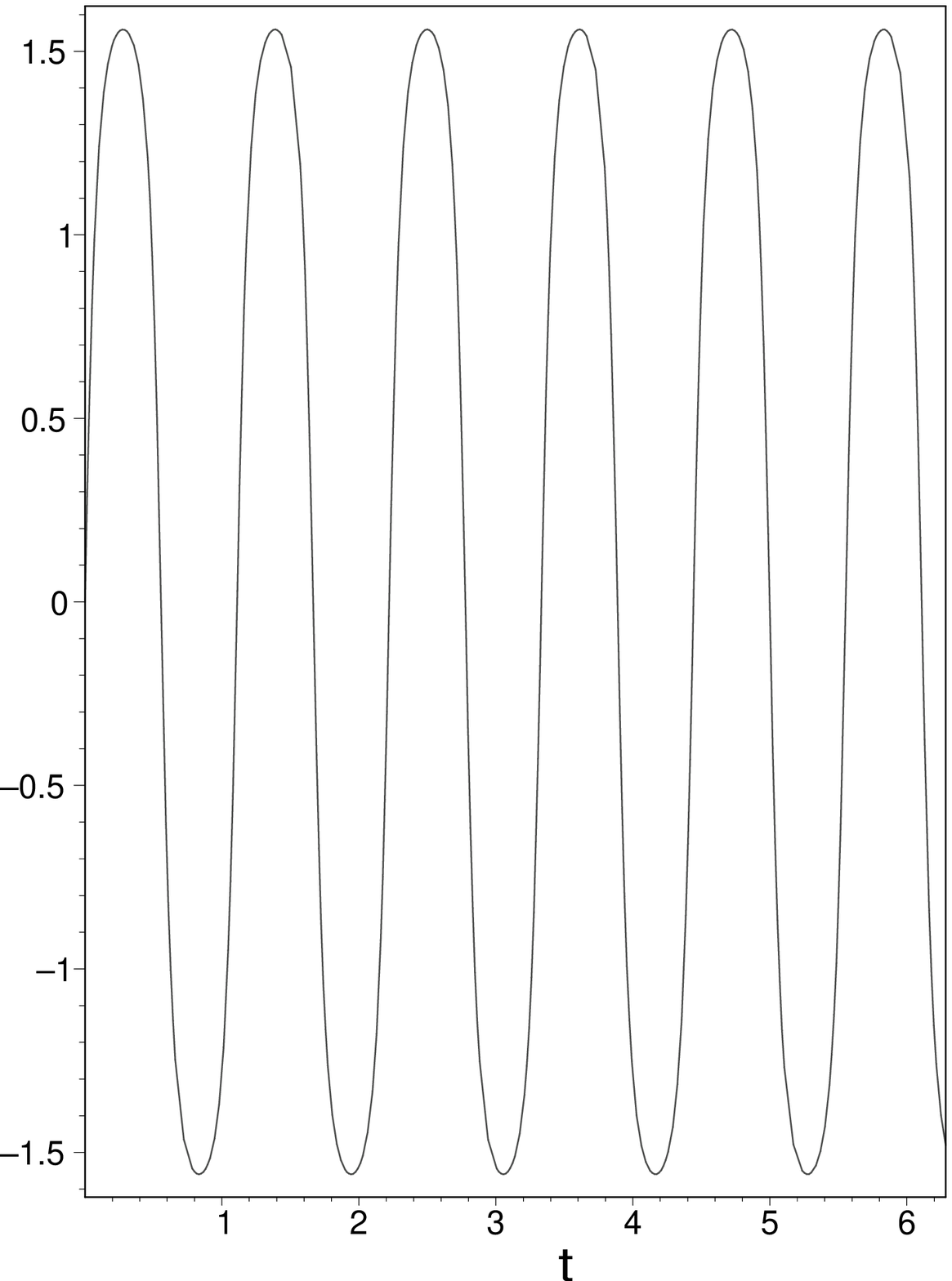}
\end{center}
\caption{The velocity $\mbox{\boldmath $v$} (t)$ of the particle in the symmetric (2,2) P\"{o}schl--Teller potential,
 for the same values of $E$ and $T$ as in  Figure \ref{figure7}
(compare  Figure \ref{figure4}).}
\label{figure8}
\end{figure}

\begin{figure}
\begin{center}
\includegraphics[width=6cm,height=6cm]{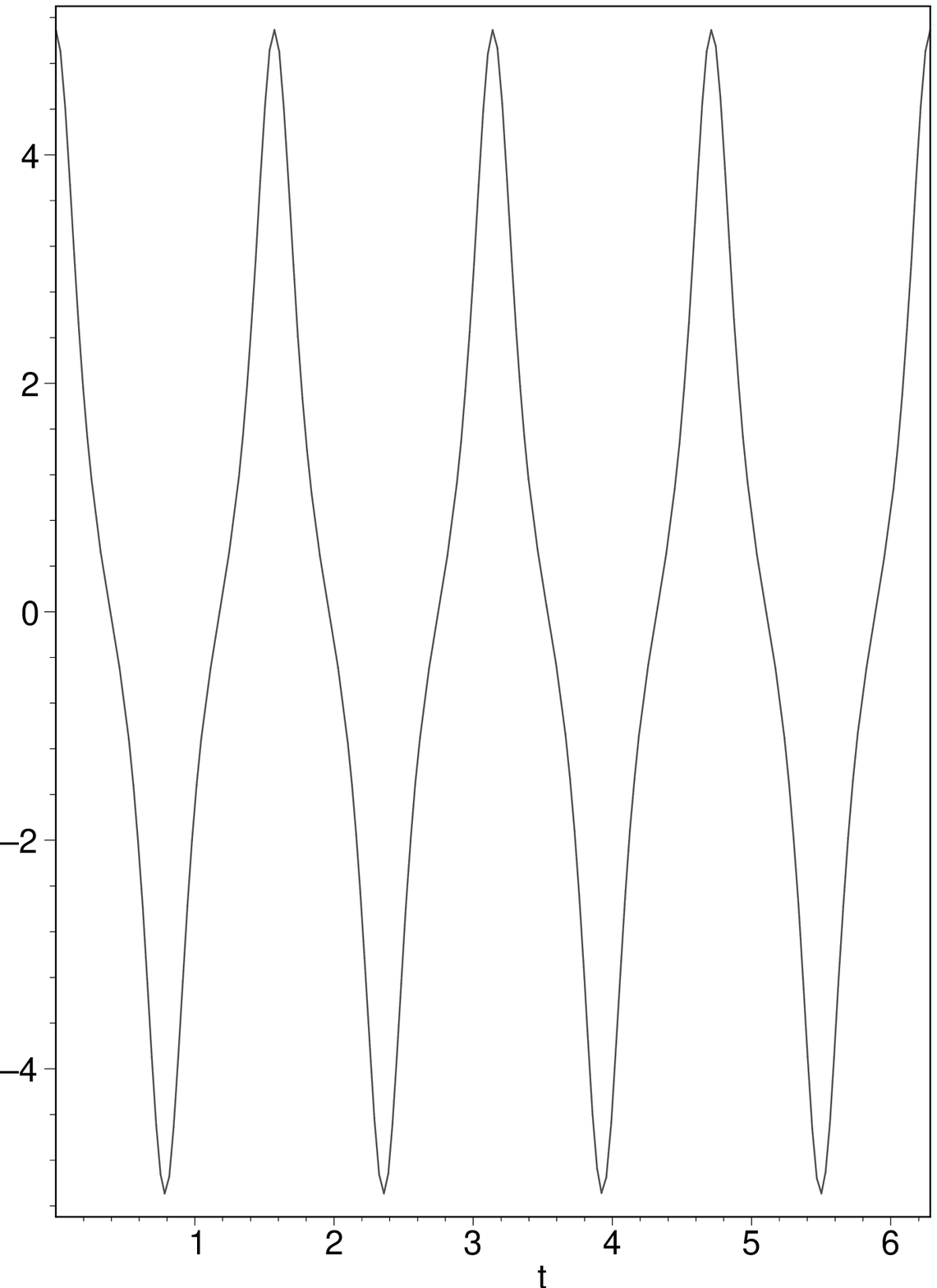}
\hspace{3mm}
\includegraphics[width=6cm,height=6cm]{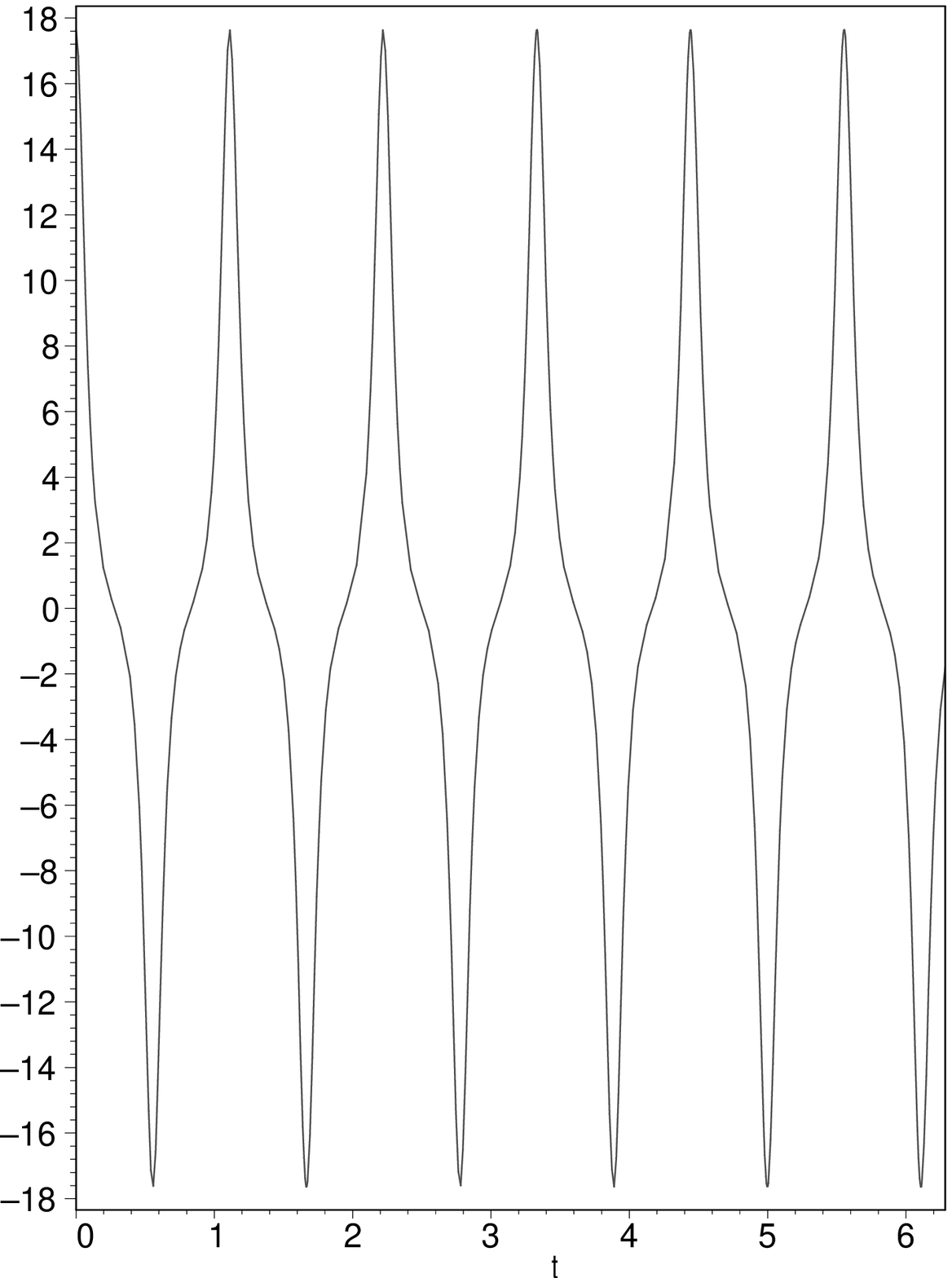}
\end{center}
\caption{The acceleration \mbox{\boldmath $\gamma$}$(t)$ of the particle in the symmetric (2,2)
P\"{o}schl--Teller potential, for the same values of $E$ and $T$ as in  Figure \ref{figure7}
(compare   Figure \ref{figure5}).} 
\label{figure9}
\end{figure}

\begin{figure}
\begin{center}
\includegraphics[width=7cm,height=7cm]{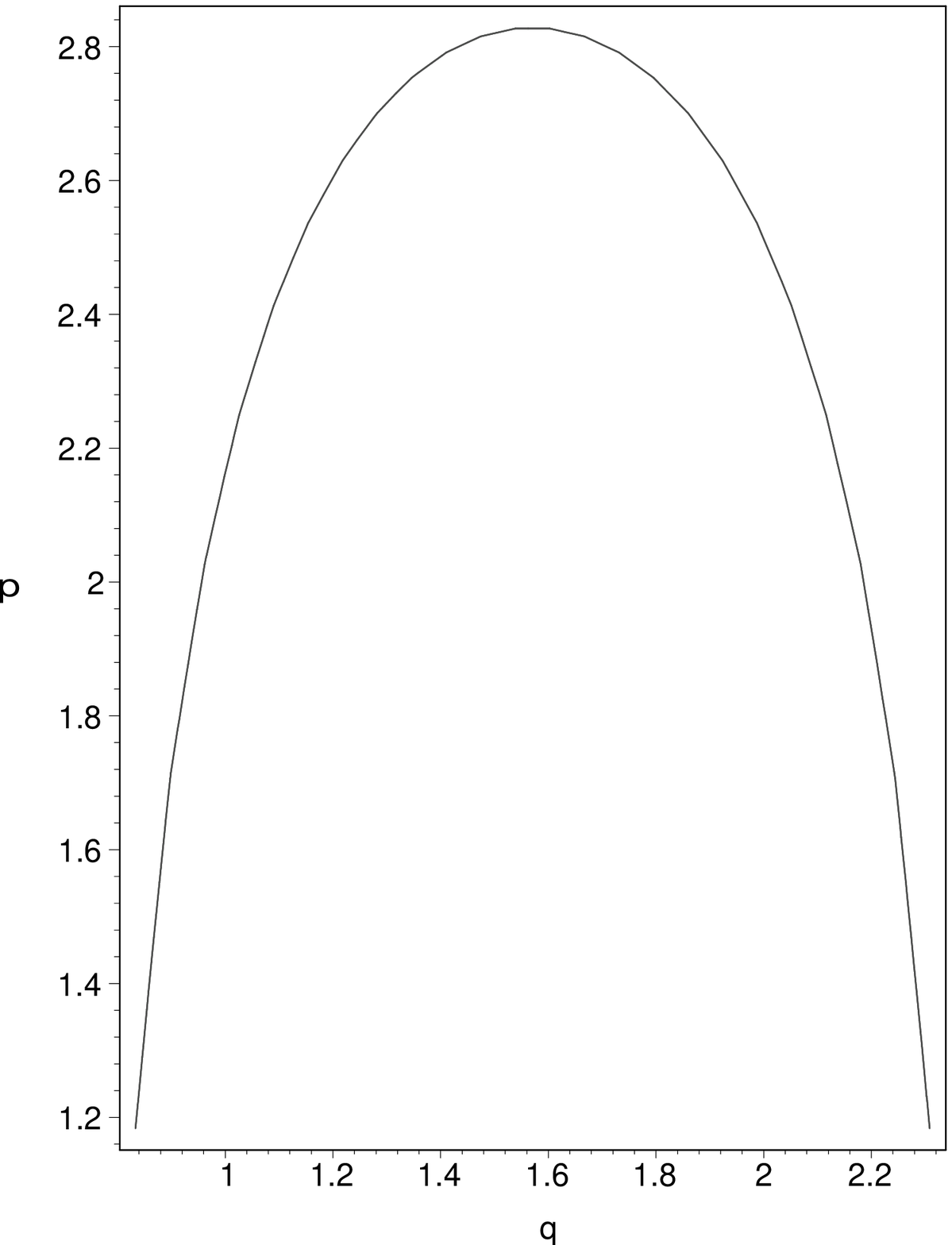}
\end{center}
\caption{Upper part of the phase trajectory of the particle  the symmetric (2,2) P\"{o}schl--Teller system,
 for the same values of $E$ and $T$ as in  Figure \ref{figure7}
(compare   Figure \ref{figure6}).} 
\label{figure10}
\end{figure}

It is worthwhile to compare (\ref{PTperiod}) and (\ref{PTaction}) with their respective infinite well counterparts 
(\ref{wellperiod}) and (\ref{wellaction}). We should also check that the time behavior (\ref{PTevol}) of $x(t)$ goes
into (\ref{wellevol}) at the limits $\alpha, \beta \to 0$. We give in Figures 
\ref{figure7}, \ref{figure8}, and \ref{figure9} the curves for   $x(t)$, $\mbox{\boldmath $v$}(t)$ and
$\mbox{\boldmath $\gamma$}(t)$, respectively, in the particular symmetric case $\lambda =\kappa = 2$, for 
two different values of the energy, namely, $E=8 V_o$ and $E=16 V_o$. Figure \ref{figure10} shows the corresponding
phase trajectory in the plane $(q = x, p = mv)$. Note that, in the general case, the equation for the latter reads
(at energy $E$):
\be
p = \pm \frac{\sqrt{2mE}}{\sin \frac{q}{a}} \left[1 - (\alpha +\beta) + (\alpha -\beta)\cos\frac{q}{a} 
- \cos^2\frac{q}{a}\right]^{1/2}. 
\en
Finally, let us give the canonical transformation leading to the action-angle variables
\bea
\varphi &=& \arccos \frac{1}{\sqrt{\Delta}}\left[ \cos\frac{q}{a} - \frac {\alpha -\beta}{2}\right] \\
A &=& a \, [ p^2 + 2mV(q) ]^{1/2}.
\ena

The Maupertuis action generating (\ref{PTaction}) is given by 
\beano
S_o(u) &=& 2 \pi nA + S_o^{princ}(u),
\\
S_o^{princ}(u)&=& - a \sqrt{2mE} \int_{u_-}^{u} [1 - (\alpha +\beta) + (\alpha -\beta)s - s^2]^{1/2} 
\frac{ds}{1-s^2}\; 
\enano
with $ u = \cos\dis\frac{x}{a}$. The last integral may be calculated explicitly, but the result is not illuminating.


\section{THE QUANTUM PROBLEM FOR THE INFINITE WELL}

Any quantum system trapped inside the infinite well $0 \leqslant x \leqslant \pi a$ must have its
wave function equal to zero outside the well. It is thus natural to impose on the wave functions the boundary conditions 
\be
\psi (x) = 0, \quad x \geqslant \pi a \quad \mbox{and} \quad x \leqslant 0. 
\label{3.1}
\en
Since the movement takes place only inside the interval $[0,\pi a]$, we may as well ignore the rest
of the line and replace the conditions (\ref{3.1}) by the following ones: \be
\psi \in L^2([0,\pi a],dx), \quad \psi (0) = \psi (\pi a) = 0. \en
Alternatively, one may consider the periodized well and impose the same periodic boundary conditions, namely,
$\psi (n\pi a) = 0, \, \forall n \in \Z$. 

In either case, stationary states of the trapped particle of mass $m$ are easily found from the eigenvalue problem 
for the Schr\"odinger operator. For reasons to be justified in the sequel, we choose the shifted Hamiltonian: 
\begin{equation} 
H \equiv H_{\rm w} = - \frac{\hbar^2}{2m} \frac{d^2}{dx^2} - \frac{\hbar^2}{2ma^2} \ .
\label{3.2}
\end{equation}
Then
\begin{equation}
\Psi (x,t) = e^{-\frac{i}{\hbar} Ht} \Psi (x,0), 
\label{time-evol} 
\end{equation}
where $\Psi (x,0) \equiv \psi (x)$ obeys the eigenvalue equation 
\begin{equation}
H \psi (x) = E \psi (x),
\end{equation}
together with the boundary conditions (\ref{3.1}). Normalized eigenstates and corresponding eigenvalues are then given by 
\bea
\psi_n (x) &=&
\sqrt{\frac{2}{\pi a}} \sin (n+1) \frac{x}{a} \; \equiv\; \langle x| n \rangle, \quad
0 \leqslant x \leqslant \pi a,
\label{PTstate}
\\
&& \nn \\
H | n \rangle &=& E_n | n\rangle , \quad n = 0,1,\ldots, \\ && \nn \\
E_n &=& \frac{\hbar^2}{2ma^2} n(n+2) \; \equiv \; \hbar \omega e_n, \ena
with
$$
\omega = \frac{\hbar}{2ma^2} \equiv \frac{2\pi}{T_r} \quad \mbox{and } \; e_n = n(n+2),
\quad n = 0,1,\ldots,
$$
where $T_r$ is the ``revival'' time to be compared with the purely classical round trip time
given in (\ref{wellperiod}).
Now the Bohr-Sommerfeld quantization rule applied to the classical action gives 
\be
a \sqrt{2mE} = A = (n+1) \hbar,
\en
so
\be
E = {(n+1)^2} \frac{\hbar^2}{2ma^2} = E_n + \frac{\hbar^2}{2ma^2}, \;n = 0,1,\ldots
\en
Thus here the Bohr-Sommerfeld quantization is exact \cite{lali},  despite the presence of the extra term
${\hbar^2}/{2ma^2}$ which follows from our particular choice of zero in the energy scale (see
(\ref{3.2})).

After these elementary considerations, let us have a closer look at the functional analysis of our problem,
following mostly \cite{simon,blank}. 
We shall denote by ${\cal H}$ the state space of the infinite well, that is, the closure of the linear span of
the orthonormal set $\{| n\rangle,\; n\in\N\}$. In the $x$-representation, of course,
$ {\cal H}= L^2([0,\pi a],dx)$.
We also denote by $AC(0,\pi a)$ the set of absolutely continuous functions on $[0,\pi a]$ whose derivatives belong
to $L^2([0,\pi a],dx)$ and by $AC^2(0,\pi a)$ the set of functions in $L^2([0,\pi a],dx)$ whose weak derivatives 
are in $AC(0,\pi a)$ (we recall that, roughly speaking, a function  is absolutely continuous iff it is the indefinite
(Lebesgue) integral of an integrable function).

We begin with the Hamiltonian (\ref{3.2}). More precisely, we define the infinite well Hamiltonian as the
unbounded operator $H \equiv {H}_{\rm w}$ in $L^2([0,\pi a],dx)$, acting as (\ref{3.2}), on the dense domain
\be
{\cal D}({H}_{\rm w}) = \{ \psi \in AC^2(0,\pi a) \mid \psi (0) = \psi (\pi a) = 0 \}.
\en
On this domain, ${H}_{\rm w}$ is self-adjoint, with purely discrete, nondegenerate spectrum
$\{ E_n = \hbar \omega e_n,\, n = 0,1,\ldots \}$, and the corresponding eigenfunctions
$\{ \psi_n,\, n = 0,1,\ldots \}$ form an orthonormal basis. Furthermore, the resolvent
$$
R_{\rm w}(-\hbar \omega) \equiv ({H}_{\rm w} + \hbar \omega)^{-1} 
= \frac{1}{\hbar \omega} \sum_{n=0}^{\infty} \frac{1}{(n+1)^2}| n\rangle \langle n |
$$
is a trace-class operator, with trace norm and Hilbert-Schmidt norm, respectively:
\beano
\|R_{\rm w}(-\hbar \omega) \|_1 &=& \frac{1}{\hbar \omega} 
\sum_{n=0}^{\infty}\frac{1}{(n+1)^2} = \frac{\pi^2}{6}\, \frac{1}{\hbar \omega} , \\
\|R_{\rm w}(-\hbar \omega) \|_2 &=& \frac{1}{\hbar \omega} 
\left[\sum_{n=0}^{\infty}\frac{1}{(n+1)^4}\right]^{1/2} = \frac{\pi^2}{\sqrt{90}}\, \frac{1}{\hbar \omega} . 
\enano 

At this stage, it is instructive to compare the Hamiltonian of the infinite well with
that of a free particle constrained on a circle of radius $a/2$. Here also, the Hilbert space is
$L^2([0,\pi a],dx)$. The Hamiltonian ${H}_{\rm c}$ has the same expression as ${H}_{\rm w} \equiv H$, but on
the domain
\be
{\cal D}(H_{\rm c}) = \{ \psi \in AC^2(0,\pi a) \mid \psi (0) = \psi (\pi a), \psi' (0) = \psi' (\pi a) \},
\en
and it is also self-adjoint on its domain. The spectrum is again purely discrete, the eigenvalues coincide
with half of those of ${H}_{\rm w}$, namely, $E_{2n-1} = \hbar \omega e_{2n-1}, \, n = 1, 2, \ldots$, 
but each of them is doubly degenerate, and there is the additional, simple eigenvalue corresponding to $n=0$, namely, 
$e_{-1} = -1$. The eigenfunctions are
\be
\left\{ \sqrt{\frac{2}{\pi a}} \sin 2n \frac{x}{a}, \; n = 1,2,\ldots; 
\sqrt{\frac{2}{\pi a}} \cos 2n \frac{x}{a}, \; n = 0,1,2,\ldots \right\}, 
\label{specHc}
\en
and they constitute another orthonormal basis of $L^2([0,\pi a],dx)$. 
Thus, there exists a unitary correspondence between the two bases (\ref{PTstate}) and (\ref{specHc}). However, the explicit form
of this map rests on the full Hilbert space structure and not only on simple trigonometric identities (see also below).

This is another instance of the well-known fact that the physics is determined by the boundary conditions,
not only by the differential expression of the operator.

Now we turn to the canonical position and momentum operators. The position operator is $Q = x$, acting on
$L^2([0,\pi a],dx)$. It is bounded and self-adjoint. As for the momentum, the natural choice is the operator
$P_o = -i d/dx$, acting on the dense domain 
\be
{\cal D}(P_o) = \{ \psi \in AC(0,\pi a) \mid \psi (0) = \psi (\pi a) = 0 \}. \label{domP}
\en
This operator is closed and symmetric, but {\em not} self-adjoint. Since its defect indices are (1,1),
$P_o$ has self-adjoint extensions, in fact an infinite number of them, indexed by the points of a unit circle, 
namely $P_\alpha = -i \hbar \, d/dx$, acting on the dense domain 
\be
{\cal D}(P_\alpha) = \{ \psi \in AC(0,\pi a) \mid \psi (\pi a) = \alpha \, \psi (0), \; |\alpha| = 1 \}.
\en
For simplicity, we choose $\alpha = 1$, that is, periodic boundary conditions. Any other choice $P_\alpha, \, \alpha \neq 1$,
 is physically acceptable, and yields similar results. 

The operator $P \equiv P_1$
is a valid candidate for the momentum observable. Its spectrum is purely discrete and nondegenerate,
$\sigma(P) = \{ 2n \hbar/a, \, n = 0, \pm1,\pm2,\ldots \}$, with corresponding eigenfunctions
$\chi_n(x) = 1/\sqrt{\pi a} \, \exp (i 2nx/a)$. The trouble is that none of these belongs to the domain of the
Hamiltonian ${H}_{\rm w}$! And indeed, one has 
\be
\frac{P^2}{2m} \neq {H}_{\rm w} + \frac{\hbar^2}{2ma^2} \ , 
\label{Psquare}
\en
since
$$
{\cal D}(P^2) = \{ \psi \in AC^2(0,\pi a) \mid \psi (0) = \psi (\pi a), \psi' (0) = \psi' (\pi a) \},
$$
so that, up to the constant ${\hbar^2}/{2ma^2}$, ${P^2}/{2m}$ coincides with the Hamiltonian ${H}_{\rm c}$ of
a particle on a circle, not ${H}_{\rm w}$! 

To conclude, we evaluate the canonical commutation relations (CCR), which take the standard form
\be
[Q,P] = i \hbar I,
\label{CCR}
\en
on the domain ${\cal D}(QP) \cap {\cal D}(PQ) = {\cal D}(P_o) $, as given in (\ref{domP}).
Correspondingly, we obtain the uncertainty relations in the eigenstates $\psi_n$ of the Hamiltonian ${H}_{\rm w}$
 [compare with the classical case, (\ref{av_pos}) and (\ref{disper})]: 
\bea
\langle Q \rangle_n	&=& \frac{\pi a}{2} \ , \nn \\ 
\langle Q^2 \rangle_n &=& a^2 \left(\frac{\pi^2}{3} - \frac{1}{2 (n+1)^2}\right), \nn \\
\langle P \rangle_n	&=& 0 \ , \\
\langle P^2 \rangle_n &=& \frac{1}{a^2} \, \hbar^2 \, (n+1)^2 \ , \nn 
\ena 
where $\langle \cdot \rangle_n \equiv \langle \psi_n \mid \cdot \mid \psi_n \rangle$.
Note that, in the last relation, $\psi_n \in{\cal D}(P)$, but $\psi_n \not \in{\cal D}(P^2)$,
so that we really mean $ \langle P^2 \rangle_n \equiv \| P\psi_n\|^2$. Also, according to   \cite{flugge},
the relation $\langle P \rangle_n= 0$ expresses the fact that the current associated to the particle vanishes
identically.

Taking all these relations together, we obtain the uncertainties 
\beano 
\langle \Delta Q \rangle_n^2 &=& \langle Q^2 \rangle_n - \langle Q \rangle_n^2 
\; = \; a^2 \left(\frac{\pi^2}{12} - \frac{1}{2 (n+1)^2}\right) \; \geqslant \; 
a^2 \left(\frac{\pi^2}{12} - \frac{1}{2}\right) , \\ 
\langle \Delta P \rangle_n^2 &=& \langle P^2 \rangle_n - \langle P \rangle_n^2
\; = \; \frac{1}{a^2} \, \hbar^2 \, (n+1)^2 \; \geqslant \; \frac{\hbar^2}{a^2} \ ,
\enano
and the uncertainty relations
\be
\langle \Delta Q \rangle_n \, \langle \Delta P \rangle_n = 
\hbar \left( \frac{(n+1)^2 \pi^2}{12} - \frac 12\right)^{1/2} \geqslant \hbar \left(\frac{\pi^2}{12} - \frac 12\right)^{1/2}
\simeq 0.57 \, \hbar \, > \frac12\, \hbar \ ,
\label{uncertrel}
\en
as expected for a quantum state which is {\em not} of minimal uncertainty. We will make similar considerations 
in Section 9 for the case of coherent states.

However, although the CCR (\ref{CCR}) look perfectly normal, they still lead to inconsistencies, because of the
unbounded character of the operators. The problem arises, for instance,  when one tries to prove the absence of
condensation in a one-dimensional interacting Bose gas \cite{bouz},  by first putting the system in a finite box of
length
$\Lambda$ with periodic boundary conditions, and then taking the thermodynamic limit $\Lambda \to \infty$. The key ingredient
 is the Bogoliubov inequality, namely \be
\frac 12 \beta\left\langle {\vphantom{^{^t}}} A A^* + A^* A \right\rangle_{\!\!\beta} \,
\left\langle{\vphantom{^{^t}}} [[C,H],C^*]\right\rangle_{\!\!\beta} \geqslant 
|\left\langle {\vphantom{^{^t}}} [C,A]\right\rangle_{\!\!\beta}|^2, \label{bogineq}
\en
where $H$ is the Hamiltonian, $\left\langle {\vphantom{^{^t}}} X \right\rangle_{\! \beta} =
{\rm Tr}\,( e^{-\beta H}X )/ {\rm Tr}\,( e^{-\beta H})$ denotes the thermal average of the observable $X$ with respect 
to the temperature $T = (k\beta)^{-1}$ and the Hamiltonian $H$. In the relation (\ref{bogineq}),
$A$ and $C$ are observables of the system which are to be chosen in a convenient way for a specific application. 
The inequality (\ref{bogineq}) is perfectly valid for bounded operators, but some care must be exercised with domains in the
case of unbounded ones, lest absurdities follow! 

In the present case, there are two possibilities. The first one \cite{bouz} consists in keeping the CCR
(\ref{CCR}), introducing a generalized notion of state as a quadratic form and generalizing the Bogoliubov
inequality (\ref{bogineq}) in a corresponding way. This indeed allows one to prove the absence of
condensation in the Bose gas for a reasonable class of interactions, including of course a gas of free particles. 

An alternative \cite{trap-lass} consists in keeping (\ref{bogineq}) unchanged, but generalizing the usual
algebraic formalism to the quasi-*algebra generated by the  operators $Q,P$. By this we mean the following.
 Define the dense domain
\be
{\cal D} = \{ \psi \in C^\infty (0,\pi a) \mid \psi^{(k)} (0) = \psi^{(k)} (\pi a), \, k = 0,1;\ldots \}.
\label{domD}
\en
Then it is easy to see that
$$
{\cal D} = \bigcap_{k=0}^\infty {\cal D}(P^k), 
$$ 
and this gives to ${\cal D}$ a natural structure of Fr\'echet space.
 From this one gets a Rigged Hilbert Space
$$
{\cal D} \subset L^2([0,\pi a],dx) \subset {\cal D}', 
$$ 
where ${\cal D}'$ denotes the strong dual of ${\cal D}$.
 Define ${\mathfrak A} = {\cal L}({\cal D},{\cal D}')$
as the space of all continuous linear maps from ${\cal D}$ into ${\cal D}'$. This space then carries a natural
structure  of quasi-*algebra in the sense of
  \cite{trapa}. Roughly speaking, this means that ${\mathfrak A}$ obeys the usual rules of algebra, except that
the product $A B$ of two elements of ${\mathfrak A}$ is well defined iff one of them leaves the domain ${\cal D}$
invariant.
 But then the canonical commutator $[Q,P]$, when viewed as an element of ${\mathfrak A}$, becomes
\be
[Q,P] = i \hbar (I - \pi a \, \hat\delta(x) ), \label{CCR_A}
\en
where $\hat\delta(x)$ denotes the multiplication operator ${\cal D} \ni \varphi \mapsto \delta(x)\varphi \in {\cal
D}'$,  an element of ${\mathfrak A}$. Then, with the modified CCR (\ref{CCR_A}), the usual Bogoliubov inequality
(\ref{bogineq}) holds on ${\mathfrak A}$ and the standard argument for proving the absence
 of condensation applies. The same reasoning can be made with any other momentum observable 
$P_\alpha, \, \alpha \neq 1$, only the r.h.s. of (\ref{CCR_A}) becomes slightly more complicated \cite{trap-lass}.

This somewhat long digression should convince the reader that the infinite well problem is really singular,
and therefore formal considerations, in particular with respect to boundary conditions, may be misleading (see, for
instance,  \cite{seki} or \cite{robinett})!

In the light of the preceding results, the time evolution (\ref{time-evol}) is trivial. On one hand, we can expand
$\Psi (x,0) \equiv \psi (x)$ in terms of the basis of eigenvectors $\psi_n$ given in (\ref{PTstate}):
$$
\Psi (x,0) = \sum_{n=0}^\infty c_n \, \,\psi_n (x), 
$$
and thus
\beano
\Psi (x,t)&=& \sum_{n=0}^\infty c_n \, e^{-\frac{i}{\hbar} Ht}\,\psi_n (x) \\ &=& \sum_{n=0}^\infty c_n \, 
e^{-\frac{i}{\hbar} E_nt}\,\psi_n (x) \\ &=& \sum_{n=0}^\infty c_n \, \psi_n (x,t) .
\enano
Alternatively, one may obtain the same result \cite{gottfried} with help of the propagator (Green function)
$G(x-x_o \, ,t)$ :
\be
\Psi (x,t) = \int_0^{\pi a} G(x-x_o \, ,t) \, \Psi (x_o,0) \,dx_o . \label{green_evol}
\en
Since the Green function is the solution with initial condition $\delta(x-x_o)$ at $t=0$ (we must take,
of course, $x_o \in (0,\pi a)$), we may write 
\beano
G(x-x_o \, ,t) &=& e^{-\frac{i}{\hbar} Ht} \,\delta(x-x_o) \\ 
&=& e^{-\frac{i}{\hbar} Ht} \,\sum_{n=0}^\infty \overline{\psi_n (x_o)} \,\psi_n (x) 
\\ 
&=& \sum_{n=0}^\infty \overline{\psi_n (x_o)} \, e^{-\frac{i}{\hbar} E_nt} \,\psi_n (x)
\\ 
&=& \sum_{n=0}^\infty \overline{\psi_n (x_o)} \, \psi_n (x,t). 
\enano 
Here we have used the relation
$$
\delta(x-x_o) = \sum_{n=0}^\infty \overline{\psi_n (x_o)} \,\psi_n (x) \qquad \Longleftrightarrow \qquad
\langle x_o \mid x \rangle = \sum_{n=0}^\infty \langle x_o \mid n \rangle \langle n \mid x \rangle,
$$
which expresses the completeness of the basis $\{ |n\rangle \}$. 

Inserting the value of $G(x-x_o \, ,t)$ into (\ref{green_evol}), we get indeed 
\beano
\Psi (x,t) &=& \int_0^{\pi a} \sum_{n=0}^\infty\overline{\psi_n (x_o)} \, \psi_n (x,t) \, \Psi (x_o \, ,0) \, dx_o\\
&=& \sum_{n=0}^\infty \psi_n (x,t) \int_0^{\pi a} \overline{\psi_n (x_o)} \, \Psi (x_o \, ,0) \,dx_o \\
&=& \sum_{n=0}^\infty c_n \, \psi_n (x,t) . 
\enano 

Next we turn to the momentum representation. Since the spectrum of the operator $P$ is discrete, the Hilbert
space in the momentum representation reduces to the space $\ell^2$ of square summable sequences. This 
is just a reformulation of the theory of Fourier series, as opposed to the Fourier integral that makes the transition between
the position and the momentum representation for a quantum mechanics on the full line $\R$.
This fact has been overlooked, for instance, in \cite{robinett}  (nothing, of course, forbids one to take the Fourier integral
transform of the infinite well wave function $\Psi (x,t)$, but the result is just a mathematically equivalent version of the
same object, {\em not} the momentum representation wave function).  Thus an arbitrary state $\psi \in
{\cal H}$ is expressed in terms of the eigenstates $\chi_n$ of $P$, 
$$ \psi = \sum_{n=-\infty}^{\infty} a_n \chi_n, \quad a_n =
\langle \chi_n \mid \psi \rangle,
\quad\mbox{thus} \quad \widehat\psi \equiv \{ a_n \} \in \ell^2. 
$$
 For instance, we obtain for the energy eigenstates 
\bea
\psi_{2k+1}(x) &=& -\frac{i}{\sqrt{2}}\left[\chi_{k+1}(x) - \chi_{-k-1}(x)\right]
\label{odd_wf} \\
\psi_{2k}(x) &=& -\frac{4}{\pi} \sum_{n=-\infty}^{\infty} \frac{2k+1}{4n^2 - (2k+1)^2} \, \chi_n (x).
\label{even_wf}
\ena
These relations constitute in fact the unitary correspondence between two different orthonormal bases, as discussed after
(\ref{specHc}), and the map is indeed nontrivial.

A last topic that would deserve to be discussed is the solution of the infinite well problem in the path
integral formalism. However this is treated in full detail in   \cite{good}, so we will refrain from reproducing
it here.


\section{THE SAME FOR P\"{O}SCHL--TELLER} 

P\"{o}schl--Teller potentials were originally introduced in a molecular physics context. The energy
eigenvalues and corresponding eigenstates are solutions to the Schr\"{o}dinger equation
\bea
\left[-\frac {\hbar^2}{2m}\, \frac{d^2}{dx^2} + \frac {V_o}{2} \left(\frac{\lambda(\lambda -1)}{\cos^2\frac{x}{2a}} 
+\frac{\kappa (\kappa -1)}{\sin^2\frac{x}{2a}}\right) -\frac{\hbar^2}{8ma^2} (\lambda+ \kappa )^2 \right]\psi (x) 
&=& E\psi (x), \nn \\
&& \makebox[-1cm]{}0\leqslant x \leqslant \pi a,
\label{schr_eqPT}
\ena
where we have also shifted the Hamiltonian of the trapped particle of mass $m$ by an amount equal to
$-\frac{\hbar^2}{8ma^2} (\lambda + \kappa)^2$. Here too, as for the infinite well, we have the choice of putting 
the potential equal to infinity outside the interval $[0,\pi a]$, or periodizing the problem, with period $2\pi a$.

Since the potential strength is overdetermined by specifying $V_o$, $\lambda$ and $\kappa$ simultaneously, we can 
freely put for convenience, as in   \cite{flugge,poschl},  
\be
V_o = \frac{\hbar^2}{4ma^2} \ .
\label{normV_o}
\en
With this choice, and the boundary conditions (BC) $\psi (0) = \psi (\pi a) = 0$, the normalized eigenstates and the
 corresponding eigenvalues, all of them simple, are given by \be
\Psi_n (x) = [c_n (\kappa, \lambda)]^{-{\frac 12}} \left(\cos \frac{x}{2a}\right)^\lambda
\left(\sin\frac{x}{2a}\right)^\kappa \;
{_2}F_1(-n,\,n+ \lambda+ \kappa \; ; \; \kappa + \frac 12 \; ; \; \sin^2 \frac{x}{2a}),
\label{eigenfPT}
\en
where $c_n (\kappa , \lambda)$ is a normalization factor that can be given analytically when $\kappa$ and
$\lambda$ are positive integers, and
\be
E_n = \frac{\hbar^2}{2ma^2}\; n(n+ \lambda + \kappa) \equiv \hbar\, \omega e_n, \; n=0,1,\ldots,
\label{hbar}
\en
with
\be
\omega = \frac{\hbar}{2ma^2},\; e_n = n(n+\lambda + \kappa), \; \lambda, \kappa >1.
\en
\medskip

Note that the Bohr-Sommerfeld rule applied to the canonical action (\ref{PTaction}) yields
(here we do {\em not} impose the normalization (\ref{normV_o})): 
$$
a \sqrt{2mE} - a \sqrt{mV_o} \left[\sqrt{\lambda (\lambda-1)} + \sqrt{\kappa (\kappa -1)}\right] = \hbar (n+\frac 12), 
$$
that is,
\bea
E_n& =& \frac{\hbar^2}{2ma^2} (n+\frac 12)^2 + \frac{\hbar}{ma} \sqrt{mV_o}(n+\frac 12)
\left[\sqrt{\lambda (\lambda-1)} + \sqrt{\kappa (\kappa -1)}\right] \nn \\ && \makebox[2.6cm]{}+ \frac {V_o}{2}
\left[\sqrt{\lambda (\lambda-1)} + \sqrt{\kappa (\kappa -1)}\right]^2. \label{PTenergy}
\ena
This formula is interesting on two counts at least. \bei
\item[(a)]
The first term in (\ref{PTenergy}) gives, apart from the term $\frac 12$ in $(n + \frac 12)$,
the exact spectrum of the infinite well. More precisely, these values of the energy may be obtained simply by letting 
$V_o \to 0$ in $V(x)$ and keeping in mind that $V = \infty$ outside $[0, \pi a ]$.

\item[(b)]
In the limit $V_o \to \infty$, the first term in (\ref{PTenergy}) can be neglected and one is left,
up to a global, $V_o$ dependent, shift,
with the spectrum of a harmonic oscillator with elementary quantum 
$$
\hbar \omega = \hbar \sqrt{\frac{V_o}{ma^2}}\left[\sqrt{\lambda (\lambda-1)} + \sqrt{\kappa (\kappa -1)}\right]. 
$$
Hence, the P\"{o}schl--Teller potential interpolates between the square-well and the harmonic oscillator.
\eni

As we did in the case of the infinite well, let us examine now the functional-analytic properties of the
P\"{o}schl--Teller Hamiltonian. The Schr\"{o}dinger equation (\ref{schr_eqPT}) is an eigenvalue equation for an 
ordinary differential Sturm-Liouville operator, which is singular at both ends of the interval $[0,\pi a]$
(see, for instance,  \cite{richt}). The situation now depends on the values of $\lambda$ and $\kappa$, as follows
from the thorough analysis of Gesztesy et al.\cite{gesz}.
In particular, there exist critical values  $\lambda,\kappa = 3/2$, although one would naively expect the value 1 
to play that role.

 Let $\dot T$ the minimal differential operator, that is,
the operator defined by the differential expression (\ref{schr_eqPT}) on the space $C_0^\infty(0, \pi a)$ of 
$C^\infty$ functions with (compact) support strictly contained in the open interval $(0, \pi a)$. Then:
\bei
\item
If $\lambda,\kappa \geqslant 3/2$, the operator $\dot T$ is in the limit point case at both ends $x = 0,\pi a$, thus 
it is essentially self-adjoint and its closure $H_{\mbox{\tiny PT}}$ automatically satisfies Dirichlet BCs at $x = 0$
and $x =\pi a$, {\em i.e.}, $\psi (0) = \psi (\pi a) = 0$.

\item
If $\lambda \geqslant 3/2 > \kappa $, the operator $\dot T$ is in the limit point case at $x = \pi a$, but in the limit 
circle case at $x = 0$; hence the defect indices of its closure are (1,1) and we need a BC at $x = 0$ for defining a
self-adjoint extension; quite naturally, we choose the Dirichlet BC, the one at $x = \pi a$ being automatic.

\item
If $\kappa \geqslant 3/2 > \lambda $, the operator $\dot T$ in the limit point case at $x = 0$, but in the limit circle 
case at $x = \pi a$; again we impose a Dirichlet BC at $x = \pi a$, the one at $x = 0$ being automatic.

\item
If $1/2 < \lambda,\kappa < 3/2$, the operator $\dot T$ is in the limit circle case at both ends $x = 0,\pi a$, the defect 
indices are (2,2), and we have to impose two BC, again chosen as Dirichlet. 

\eni
Notice that the Dirichlet BC may be written as $\psi (0) = \psi (\pi a) = 0$ in the first, regular, case, but it takes 
a more complicated form in the singular cases \cite{gesz}.  Clearly this choice of boundary conditions is dictated by
physics, namely, it is the same as for the infinite well. One may also say that the chosen self-adjoint extension of 
$\dot T$ is obtained by analytic extension from large positive values $\lambda,\kappa \geqslant 3/2$, since, in this context,
everything depends analytically on $\lambda,\kappa$. 

In all four cases, we define the P\"{o}schl--Teller Hamiltonian as the self-adjoint operator $H_{\mbox{\tiny PT}}$ 
in $L^2([0,\pi a],dx)$, acting as the left-hand side of (\ref{schr_eqPT}), on the dense domain
\bea
{\cal D}({H}_{\mbox{\tiny PT}})& =& \{ \psi \in AC^2(0,\pi a) \mid V_{\mbox{\tiny PT}}\psi \in L^2([0,\pi a],dx) 
\mbox{ and $\psi$ satisfies } \nn
\\
&&\makebox[5cm]{} \mbox{ a Dirichlet BC at } x = 0,\pi a \}, 
\ena
where $V_{\mbox{\tiny PT}}$ is the P\"{o}schl--Teller potential. The P\"{o}schl--Teller Hamiltonian has pure point
spectrum,  without multiplicity, and given by $E_n = \frac{\hbar^2}{2ma^2}\; n(n+ \lambda + \kappa) \equiv \hbar\,
\omega e_n, \; n=0,1,\ldots$, as given in (\ref{hbar}), with corresponding eigenvectors (\ref{eigenfPT}). Notice that
these eigenfunctions belong to the domain ${\cal D}({H}_{\mbox{\tiny PT}})$, since they satisfy the boundary
conditions (by assumption \cite{poschl}). 

Several remarks are in order at this point. 

(1) First, the case $ 1/2 < \lambda, \kappa < 1$  (the mixed cases have no physical relevance) corresponds to the inverted
well, yet the spectrum of ${H}_{\mbox{\tiny PT}}$ remains unchanged, that is, pure point and positive. Although the
potential is now attractive, it is too close to the walls to allow negative energy bound states. This counterintuitive
situation follows, of course, from the Dirichlet BC that make the walls impenetrable and thus confine the particle inside of
the interval. On the other hand, for $ \lambda$ or $ \kappa < 1/2$, the problem is of a different nature and the analysis of
  \cite{gesz} does not apply any more (presumably, here one faces again the ``fall towards the center'' 
phenomenon~\cite{lali}).

 (2) Next, one may choose different BC for defining a self-adjoint extension of $\dot T$. An interesting choice is to
take  the full periodicity interval $[-\pi a,\pi a]$, that is, $(-\pi a,0) \cup (0,\pi a)$, and to impose to both
$\psi$ and $\psi'$ continuity conditions at $x = 0$ and periodic BC at $x = \pm\pi a$. The resulting self-adjoint 
Hamiltonian ${H}_{\mbox{\tiny PT}}^{\rm per}$ also has a pure point spectrum, namely $\{n(n+\lambda + \kappa),
(n+1)(n+1-\lambda - \kappa), \, n = 0,1,2,\ldots\}$, with all eigenvalues simple. For $\lambda = \kappa = 1$, one
indeed recovers the doubly degenerate spectrum of the circle Hamiltonian ${H}_{\rm c}$ of Section III. 

(3) Finally, the real difference with respect to the values of $\lambda,\kappa$ comes when one periodizes the
P\"{o}schl--Teller Hamiltonian over the whole line, that is, on
$\R \setminus \pi a \Z$, with continuity BCs at $x \in \pi a \Z$. Then, if $\lambda \geqslant 3/2$ or
$\kappa\geqslant 3/2$,  the periodized Hamiltonian ${H}_{\R}^{\rm per}$ is self-adjoint, and has a pure point
spectrum, with each eigenvalue of infinite multiplicity. On the contrary, if $1/2 < \lambda, \kappa \leqslant 3/2$,
then ${H}_{\R}^{\rm per}$ really looks as the Hamiltonian of a 1-D crystal,  and indeed, it has no eigenvalue and its
spectrum has a band structure, that is, it is purely continuous with infinitely many gaps \cite{scarf,li-kus,gesz}. 
\medskip

Coming back to the interval $[0,\pi a]$, the resolvent of the P\"{o}schl--Teller Hamiltonian $H_{\mbox{\tiny PT}}$
reads 
$$
R_{\mbox{\tiny PT}}(-\frac14 \hbar \omega (\lambda+\kappa)^2) \equiv ({H}_{\mbox{\tiny PT}}
+ \frac14 \hbar \omega (\lambda+\kappa)^2)^{-1} = \frac{1}{\hbar \omega}
 \sum_{n=0}^{\infty} \frac{1}{[n+\frac12(\lambda+\kappa)]^2}
| n,\lambda,\kappa\rangle \langle n,\lambda,\kappa |, 
$$
where $| n,\lambda,\kappa\rangle$ denotes the eigenfunction $\Psi_n$ of (\ref{eigenfPT}).
As before, it is a trace-class operator, with trace norm: 
$$
\|R_{\mbox{\tiny PT}}(-\frac14 \hbar \omega (\lambda+\kappa)^2) \|_1 = \frac{1}{\hbar \omega}
\sum_{n=0}^{\infty}\frac{1}{[n+\frac12(\lambda+\kappa)]^2} \ . 
$$
Note that the Hilbert space and the momentum observable $P$ remain the same as in the case of the infinite well. 
Thus, the previous discussion remains valid and the same difficulties are present. For instance, as for 
(\ref{Psquare}),
$P^2/2m$ does {\em not} coincide with the first term of ${H}_{\mbox{\tiny PT}}$. Also one can calculate, at least in
principle,  the analogues of (\ref{odd_wf}) and (\ref{even_wf}), which are simply the Fourier series expansion of the
P\"{o}schl--Teller energy eigenstates $\Psi_n $. 


\section{THE LIMITS}

In this section, we shall investigate various limiting cases. Let us begin with the infinite square-well.
Since the natural dimensionless variable is $y = x/a$, we may rewrite the Hamiltonian (\ref{3.2}) in terms of $y$, 
and we get the scaling equation
\be
H_{\rm w} \equiv H_{\rm w}[a] = \frac{1}{a^2} \, H_{\rm w}[1]. \en The operator $H_{\rm w}[1]$ is self-adjoint 
in $L^2([0,\pi], dy)$, its eigenvalues are
$E_n[1] = \frac{\hbar^2}{2m} n(n+2)$, and one has the scaling law $E_n[a] = \frac{1}{a^2}\, E_n[1]$. From these
relations, the two limits $a \to 0$ (infinitely narrow well) and $a \to \infty$ (infinitely large well)
are trivial. The spectrum keeps the same shape, only the eigenvalues scale as $1/a^2$.

The same considerations apply to the P\"{o}schl--Teller Schr\"{o}dinger equation (\ref{schr_eqPT}). If we do not 
impose the normalization relation (\ref{normV_o}), we get the scaling relation \be
H_{\mbox{\tiny PT}} \equiv H_{\mbox{\tiny PT}}[a,V_o] = \frac{1}{a^2} \, H_{\mbox{\tiny PT}}[1,a^2 V_o].
\en
With (\ref{normV_o}), this becomes, exactly as for the infinite well, \be
H_{\mbox{\tiny PT}}[a] = \frac{1}{a^2} \, H_{\mbox{\tiny PT}}[1], \en
and the same for the eigenvalues. Thus, when the well gets narrower as $a \to 0$, the eigenvalues increase as $1/a^2$, 
but the spectrum keeps the same shape. Similarly, $a \to \infty$ implies $V_o \to 0$, and the spacing between successive 
eigenvalues goes to zero: in the limit, we recover a free particle, with continuous energy spectrum $[0,\infty]$.

Next we analyze the limit $\lambda, \kappa \to 1$, that is, the limit P\"{o}schl--Teller $\to$ infinite square-well.
For simplicity, we take the symmetric case $\lambda = \kappa$, with P\"{o}schl--Teller potential (\ref{symmPT}). 
As in the case of the infinite square-well, the symmetric P\"{o}schl--Teller Hamiltonian is self-adjoint and its resolvent
$$
R_{\mbox{\tiny PT}}(-\hbar \omega\lambda^2) \equiv ({H}_{\mbox{\tiny PT}} + \hbar \omega \lambda^2)^{-1} =
\frac{1}{\hbar \omega} \sum_{n=0}^{\infty} \frac{1}{(n+\lambda)^2}| n,\lambda\rangle \langle n,\lambda |
$$
is a trace-class operator, with trace norm: 
$$
\|R_{\mbox{\tiny PT}}(-\hbar \omega \lambda^2) \|_1 = \frac{1}{\hbar \omega} \sum_{n=0}^{\infty}\frac{1}{(n+\lambda)^2} \ . 
$$

As for the limit P\"{o}schl--Teller $\to$ infinite square-well, the exact statement is that
${H}_{\mbox{\tiny PT}} \to H_{\rm w}$ in strong resolvent sense as $\lambda \to 1$, that is,
$R_{\mbox{\tiny PT}}(z) \to R_{\rm w}(z)$ strongly, for all nonreal $z$. This follows from 
  \cite{simon}, Theorem VIII.25 (a), 
as we now prove. 
The domain $C_0^\infty(0, \pi a)$ is dense in $L^2([0, \pi a],dx)$ and it is a core both for
$H_{\rm w}$ and for $ H_{\mbox{\tiny PT}}[\lambda]$, for any $\lambda \geq 3/2$ \cite{blank}.  Then, we obtain a core
for $ H_{\mbox{\tiny PT}}[\lambda], \; 1 <\lambda < 3/2$, by taking the set
${\cal D}_{\mbox{\tiny PT}} = \{\psi = \phi + c_1 \psi_1 + c_2 \psi_2, \, \phi \in C_0^\infty(0, \pi a)\}$, where
$\psi_1$  and $\psi_2$ are two solutions of $H_{\mbox{\tiny PT}}\chi = k^2 \chi, \; {\rm Im}\, k \geq 0 $, chosen in
such a way that
$\psi$ obeys the boundary conditions that define $H_{\mbox{\tiny PT}}$. In our case of Dirichlet BC, this implies
\cite{gesz} that $c_1=0$ and $\psi_2 = \Psi_n$, the eigenfunction (\ref{eigenfPT}), taken for $\lambda = \kappa$.

Choose any decreasing sequence $\{\lambda_j, \, j=1,2,\ldots; \lambda_j >1, \lambda_j \to 1
\mbox{ as } j \to \infty\}$. Then, $ H_{\mbox{\tiny PT}}[\lambda_j] \psi \to H_{\rm w}\psi$,
for each $\psi = \phi + c \psi_2 \in {\cal D}_{\mbox{\tiny PT}}$. Indeed, \jot3mm
\beano
\| H_{\mbox{\tiny PT}}[\lambda_j] \psi - H_{\rm w}\psi \| &=& \left\|V_{\lambda_j}(x) \psi - 
\frac{\hbar^2}{2ma^2}(\lambda_j^2 - 1)\psi \right\| \\
&=& \frac{\hbar^2}{ma^2}\left\| \lambda_j(\lambda_j-1)\left[\sin \frac{x}{a}\right]^{-2}\,\psi
-\frac12(\lambda_j^2 -1)\psi \right\|	\\ &=& (\lambda_j-1)\frac{\hbar^2}{ma^2}\left(\lambda_j\left\| 
\left[\sin \frac{x}{a}\right]^{-2}\,\psi \right\|+
\frac12(\lambda_j +1) \|\psi \|\right) \\ &\to& 
0 \mbox{ as } j \to \infty,
\enano
\jot0mm
since both $\phi \in C_0^\infty(0, \pi a)$ and $\psi_2 = \Psi_n$ belong to the domain of $[\sin \frac{x}{a}]^{-2}$. 
By the theorem quoted, this implies that ${H}_{\mbox{\tiny PT}} \to H_{\rm w}$ in strong resolvent sense. As a
consequence,
 by   \cite{simon}, Theorem VIII.24, for each eigenvalue $E_n = \hbar\, \omega\; n(n+ 2)$ of the limiting
operator
$H_{\rm w}$, there is, for each $j$, an eigenvalue $E_n[\lambda_j] = \hbar\, \omega\; n(n+ 2\lambda_j)$ of
$H_{\mbox{\tiny PT}}[\lambda_j]$ such that
$E_n[\lambda_j] \to E_n$ as $ j \to \infty$. Put in a simpler way, the eigenvalues $E_n[\lambda] $ are continuous in
$\lambda$ and $E_n[\lambda] \to E_n$ as $\lambda \to 1$, for each $n= 0,1,2,\ldots$..

Finally, there is the semiclassical limit $ \hbar \to 0, \; n\hbar = $ const, but this problem is
fully treated in the literature, for instance, in   \cite{lali}, so we omit it.


\section{THE DYNAMICAL ALGEBRA ${\mathfrak{su}}(1,1)$}

 Behind the spectral structure of the infinite well
 or P\"{o}schl--Teller Hamiltonians, there exists a dynamical algebra generated by lowering and raising operators
  \cite{ga-mon,gaz-champ}.  The latter are defined by
\bea
a | n \rangle &=& \sqrt{e_n} | n-1 \rangle \label{5.1}\\ 
a\da | n \rangle &=& \sqrt{e_{n+1}} | n+1 \rangle
\label{5.2}
\ena
with
\beano
e_n &=& n(n+2), \; \mbox{for the infinite well,} \\ e_n &=& n(n+\lambda + \kappa), \; \mbox{for the
P\"{o}schl--Teller potential,} \; n=0,1,2,\ldots \enano
Then we note that the operator
\begin{equation}
X_N = a\da a
\end{equation}
is diagonal with eigenvalues $e_n$:
\begin{equation}
X_N | n\rangle = e_n | n\rangle \ .
\end{equation}
Note that the number operator $N$,
\begin{equation}
N | n\rangle = n | n\rangle \ ,
\end{equation}
is given in terms of $X_N$ by
\begin{equation}
N = - \frac12(\lambda + \kappa)+   \left( X_N +   \frac14 (\lambda + \kappa)^2\right)^{1/2} \ . 
\end{equation} 
For any diagonal operator $\Delta$ with
eigenvalues $\delta_n$ \be \Delta |n \rangle = \delta_n | n\rangle, \en we denote its finite difference by $\Delta'$. 
The latter is defined as the diagonal operator with eigenvalues $\delta'_n \equiv \delta_{n+1}-\delta_n$,
\be
\Delta' | n\rangle = \delta'_n | n\rangle \en
More generally, the $m$th finite difference $\Delta^{(m)}$ will be recursively defined by
\be
\Delta^{(m)} = (\Delta^{(m-1)})'.
\en
Now, from the infinite matrix representation (in the basis ${|n \rangle})$ of the operators
$a$ and $a\da$,
\bea
a &=& \left(
\begin{array}{ccccc}
0 & \sqrt{e_1} & 0 & 0 &\ldots\\
0 & 0 & \sqrt{e_2} & 0 &\ldots \\
0 & 0 & 0 &\sqrt{e_3} &\ldots \\
\ldots &\ldots &\ldots &\ldots & \ldots
\end{array}
\right) ,
\\
a\da &=& \left(
\begin{array}{ccccc}
0 & 0 & 0 & 0& \ldots\\
\sqrt{e_1} & 0 & 0 & 0 &\ldots\\
0 & \sqrt {e_2} & 0 & 0 & \ldots\\
0 & 0 & \sqrt{e_3} & 0 & \ldots \\
\ldots &\ldots &\ldots &\ldots & \ldots
\end{array}
\right)
\ena
it is easy to check that
\be
{[a,a\da ]} = \left(
\begin{array}{cccc}
e_1 - e_0 & 0 &\ldots\\
0 & e_2 - e_1 &\ldots & 0 \\
0 & 0 & e_3 - e_2
\end{array}
\right)
= X'_N
\en
\be
X'_N | n\rangle = e'_n | n\rangle,\quad e'_n = e_{n+1} - e_n = 2n+3, \mbox{ resp. } 2n+1+\lambda+\kappa.
\label{5.13}
\en
We also check that, for any diagonal operator $\Delta$, we have \be \begin{array}{rcl}
{[a, \Delta]} &=& \Delta' a \ ,\\
{[a\da, \Delta]} &=& - a\da \Delta' \ .
\end{array}
\en
Therefore,
$$
[a, X'_N] = X''_N a,
$$
with
\be
X''_N | n\rangle = e''_n | n\rangle = (e'_{n+1} - e'_n) | n\rangle = 2 | n\rangle.
\en
So
\begin{equation}
X''_N = 2 I, \quad X'''_N = 0 \ ,
\end{equation}
and
\be
[a, X'_N] = 2a.
\en
Similarly,
\be
[a\da, X'_N] = - 2a\da.
\en
In summary, there exists a ``dynamical'' Lie algebra, which is generated by $\{a, a\da, X'_N\}$.
Then the commutation rules
\be 
{[a, a\da]}= X'_N, \quad
{[a, X'_N]} = 2a, \quad
{[a\da, X'_N]} = - 2a\da,  
\en 
clearly indicate that it is isomorphic to 
\be 
\mathfrak{su}(1,1) \sim \mathfrak{ sl} (2,\R) \sim \mathfrak {so} (2,1) . 
\label{lie-alg}
\en
A more familiar basis for (\ref{lie-alg}) is given (in $\mathfrak {so}(2,1)$ notation) by
\begin{equation}
L^- = \frac{1}{\sqrt2} a , \quad L^+ = \frac{1}{\sqrt2} a\da, \quad L_{12} = \frac12 X'_N,
\label{5.21}
\end{equation}
where $L_{12}$ is the generator of the compact subgroup $SO(2)$,namely,
\be 
{[L^{\pm}, L_{12}]} = \mp L^{\pm},  \quad {[L^- , L^+]} = L_{12}.
\en
Note that if we add the operator $X_N$ ( {\em i.e.}, the Hamiltonian $H$) to the set $\{a, a\da, X'_N\}$,
we obtain an infinite-dimensional Lie algebra contained in the enveloping algebra.
Indeed
\begin{equation}
\begin{array}{llllll}
{[a, X_N]} &=& X'_N a, &\quad {[a\da, X_N]} &=& - a\da X'_N \\ {[a, X'_N a]}
&=& 2a^2, &\quad {[a\da, X'_N a]}
&=& - {X'_N}^2 - 2 X_N, \\
&& \mbox{etc}\ldots &&&
\end{array}
\end{equation}
Note also the relation between $X_N$ and $X'_N$: 
\begin{equation}
X_N = \frac14 ({X'_N}^2 - 2 X'_N -3), \mbox{ resp. } \frac14 ({X'_N}^2 - 2 X'_N -(\lambda+\kappa+1)(\lambda+\kappa)). 
\end{equation}
In the same vein, we note that the condition $X'''_N = 0 $ is necessary for obtaining a genuine Lie algebra (instead
of a subset of the enveloping algebra). Therefore, $\mathfrak{su}(1,1)$ is the {\em only} dynamical Lie algebra that
can arise in such a problem.

It follows from the considerations above that the space $\cal H$ of states $ | n \rangle$ carries
some representation of
$\mathfrak{su}(1,1)$. The latter is found by examining the formulas for the $\mathfrak{su}(1,1)$ discrete series 
representation \cite{vil,gelf-vil,knapp}.

Given $\eta = \frac12, 1, \frac32,\ldots$ the discrete series UIR $U_\eta$ is realized on the Hilbert
space ${\cal H}_\eta$ with basis $\{| \eta, n \rangle, n \in \N\}$ through the following actions of the
Lie algebra elements
\bea
L_{12} | \eta, n\rangle &=& (\eta + n) \; | \eta, n\rangle \label{6.26}\\ L^- | \eta, n \rangle
&=& \frac{1}{\sqrt2} \sqrt{(2\eta +n-1)n} \;| \eta, n-1\rangle \label{6.27}\\ L^+ | \eta,n\rangle &=& \frac{1}{\sqrt2} 
\sqrt{(2\eta +n)(n+1)} \;| \eta, n+1\rangle
\label{6.28b}
\ena
The representation $U_\eta$ fixes the Casimir operator \begin{equation} Q = - L_{12} (L_{12} -1) + 2 L^+ L_-
\end{equation}
to the following value
\begin{equation}
Q {\cal H}_\eta = \eta (\eta-1) {\cal H}_\eta\ . 
\end{equation} 
Using (\ref{5.21}) and (\ref{6.27}), and
comparing with (\ref{5.1}), (\ref{5.2}), (\ref{5.13}), we obtain the specific value of $\eta$ for 
the infinite well problem, namely, $\eta = \frac32$, so that we can make the identifications 
${\cal H}_{3/2} \equiv {\cal H} , \; | \frac32, n\rangle \equiv | n\rangle$.
On the other hand, we obtain a continuous range of values for the P\"{o}schl--Teller potentials:
\be
\eta = \frac{\lambda+\kappa+1}{2} > \frac32 \ , \label{PT_eta} 
\en 
and we shall denote the corresponding Hilbert spaces and states (\ref{PTstate}) by ${\cal H}_\eta$ and
$ | \eta, n \rangle$, respectively. The relation (\ref{PT_eta}) simply means that we are  here  in presence of
  the (abusively called) discrete series representations of the universal covering of $SU(1,1)$, except for the
interval  $\eta \in (\frac 12, \frac 32).$


\section{COHERENT STATES FOR THE INFINITE WELL}

In a general setting, consider a strictly increasing sequence of positive numbers
\be
0 = e_0 < e_1 < e_2 \ldots < e_n < \ldots, \label{numbers} \en which are eigenvalues of a self-adjoint positive operator
 $X_N$ in some Hilbert space $\cal H$,
\be
X_N | n\rangle = e_n | n\rangle,
\en
where the set $\{| n\rangle, \; n\in \N \}$ is an orthonormal basis of $\cal H$ \cite{ga-kl,gaz-champ}.

There corresponds to (\ref{numbers}) a (generically infinite) dynamical Lie algebra with basis
$\{a,a\da, X'_N, \ldots \}$, with the notation of the previous section. There also corresponds a
continuous family $\{| z\rangle, \; z \in C(0,R) \subset \C\}$ ($C(0,R)$ is the open disk of center 0 and radius $R$) 
of normalized coherent states, eigenvectors of the operator $a$: \be a| z \rangle = z | z \rangle.
\en
The explicit form of those coherent states is \begin{equation}
| z \rangle = \frac{1}{N (| z |^2)} \sum_{n \geqslant 0} \frac {z^n}{\sqrt{\rho_n}}\; | n\rangle,
\label{cohst}
\end{equation}
where
\be
\rho_0 = 1, \quad \rho_n = e_1 \; e_2 \; \ldots \; e_n, \; n > 0. \label{rho_n_well}
\en
$N(| z|^2)$ is a normalization factor:
\begin{equation}
\langle z | z \rangle = 1 \quad \Longleftrightarrow \quad \left(N(| z |^2 )\right)^2
= \sum_{n=0}^{+ \infty} \frac{| z |^{2n}}{\rho_n}. \end{equation}
Of course, these coherent states exist only if the radius of convergence 
\begin{equation}
R = \limsup_{n\to +\infty} \sqrt[n]{\rho_n} \end{equation}
is nonzero. 
In fact, different specific choices of $\rho_n$ give rise to many different families of coherent states, 
as illustrated in a series of recent works \cite{penson,sixden1,sixden2,klau-penson}.

Now suppose that $X_N$ is (up to a factor) the Hamiltonian for a quantum system,
\begin{equation}
H = \hbar \omega X_N.
\end{equation}
Then the coherent states (\ref{cohst}) evolve in time as \be
e^{-\frac{i}{\hbar}Ht} \;| z\rangle = \frac {1}{N(| z |^2)} \sum_{n\geqslant 0}\frac{z^n}{\sqrt{\rho_n}}\;
 e^{-i\omega e_{n}t}\; | n \rangle.
\en
If $e_n \propto n$,  {\em i.e.}, in the case of the harmonic oscillator, the temporal evolution of the 
coherent state $| z\rangle$ reduces to a rotation in the complex plane, namely,
$e^{- iHt/\hbar} \;| z\rangle = | z\, e^{-i\omega t} \rangle$.
In general, however, we will lose the temporal stability of our family of coherent states (\ref{cohst}).
Hence, in order to restore it, we must extend our original definitions 
to the entire orbits 
\begin{equation}
\{e^{-i \frac {H}{\hbar} t}\; | z \rangle , \; z \in C(0,R), \; t \in I \}. \end{equation}
The interval $I$ is the whole real line when $e_n$ is generic, whereas it can be restricted
to a period, that is, a finite interval $[a,b]$, \begin{equation} b-a = \frac {2\pi}{\omega \alpha}
\end{equation}
if $e_n \in \alpha \N$. A straightforward calculation now shows that \be \langle z | H | z \rangle 
= \langle z | \hbar \omega X_N | z\rangle = \hbar \omega | z |^2.
\label{PTmeanenergy}
\en
Therefore the quantity $| z |^2$ is the average energy evaluated in the elementary quantum unit $\hbar\omega$.
Note that \be
\hbar| z |^2 \equiv J
\en
is simply the action variable in the case where $H$ is the Hamiltonian of the harmonic oscillator and
the variable $z$ is given the meaning of a classical state in the phase space $\C$.
Indeed, the given choice of $\rho_n$ in (\ref{rho_n_well}) ensures that $\langle z|H|z\rangle =\omega J$
for a general Hamiltonian.

On the other hand, introducing the dimensionless number \begin{equation} \gamma = \omega t, \; \gamma \in \omega I, 
\end{equation} we are naturally led to study the continuous family of states \begin{equation} | z, \gamma \rangle =
\frac{1}{N(J)}
\sum_{n \geqslant 0} \frac {z^{n}\,e^{-i\gamma e_n}}{\sqrt{\rho_n}} \;| n\rangle .
\end{equation}
These states, parametrized by $(z, \gamma) \in C(0,R) \times I$, may be called ``coherent''
for several reasons. First they are, by construction, eigenvectors of the operator
\be
a (\gamma ) \equiv e^{-i\gamma H/\hbar \omega} \; a \; e^{i\gamma H/\hbar \omega},
\en
namely,
\be
a (\gamma) | z, \gamma \rangle = z | z, \gamma \rangle. \en They obey the temporal stability condition 
\be 
e^{-i Ht /\hbar}\, | z, \gamma \rangle = | z, \gamma + \omega t \rangle. 
\en 
Again, if we consider the harmonic oscillator
case, we do not make any distinction between the argument of the complex parameter $z$ and the angle variable $\gamma$, since
then $e_n = n$ and
$z^n e^{-i \gamma n}= (z e^{-i \gamma} )^n \ , $ so that the only parameters we need are $J = |z|^{1/2}$ and $\gamma$.
 The latter are easily identified with
the classical action-angle variables. We shall stick to the minimal parametrization set in the present generalization 
and shall denote from now on our coherent states by \be
| J, \gamma \rangle = \frac{1}{N(J)}
\sum_{n \geqslant 0} \frac {J^{n/2}\,e^{-i\gamma e_n}}{\sqrt{\rho_n}} \;| n\rangle .
\label{cohst-well}
\en
In a suitable way \cite{ga-kl} (see also the discussion in Section X), it is also acceptable to regard the
parametrization
$(J,\gamma)$ as ``action-angle'' variables, and it is convenient to refer to them as such, even when
keeping in mind the possibility of extending $\sqrt{J}$ to the complex plane,  {\em i.e.}, replacing $\sqrt{J}$ by $z$.

Let us now make those things explicit in our problem of the infinite well. In that case,
\be
\rho_n = e_1 \; e_2\; \ldots \; e_n = \frac {n! (n+2) !}{2}, \label{rho} 
\en
\be
| J, \gamma \rangle = \frac {1}{N(J)}
\sum _{n\geqslant 0} \frac{J^{n/2}\, e^{-i\gamma n(n+2)}}{\sqrt{\frac{n!(n+2)!}{2}}} | n \rangle. \label{jgamma}
\en
The normalization factor is easily calculated in terms of the modified Bessel function $I_\nu$ \cite{magnus-ob}.
\bea
(N(J))^2 &=& 2 \sum _{n=0}^{+\infty} \frac {J^n}{n!(n+2)!} \\ 
&=& \frac{2}{J} I_2 (2\sqrt J).
\ena
The radius of convergence
$R = \limsup_{n \to + \infty} \sqrt [n] {\frac{n!(n+2)!}{2}}$ is of course infinite.
Moreover, since the $e_n$'s are here natural numbers, the interval of variation of the evolution parameter
$\gamma$ can be chosen as $I = [0, 2\pi ] $. 

The positive constants $\rho_n$ arise as moments of a probability distribution $\rho(u)$,
\begin{equation}
\rho_n = \int_0^{\infty} u^{n} \, \rho (u) \,du. \label{rho_n}
\end{equation}
Also, $\rho (u)$ is explicitly given in terms 
of the other modified Bessel function $K_\nu$ \cite{magnus-ob},
 \begin{equation}
\rho (u) = u K_2 (2 \sqrt{u} )
\label{rhomeas}.
\end{equation}
It is then immediate to check that the family $\{ | J, \gamma \rangle , \; J \in \R^+, \gamma \in [0 , 2\pi ] \}$
resolves the unit operator,  {\em i.e.},
\begin{equation}
I = \int | J, \gamma \rangle \langle J, \gamma | \; d\mu (J, \gamma ),
\end{equation}
with
\begin{equation}
\int (\cdot) \,d\mu (J,\gamma )
 = \frac {1}{2\pi} \int_{-\pi}^{\pi} d\gamma \int_0^{+\infty} k(J) (\cdot) \, dJ ,
\end{equation}
where
\bea
k(J) &=& N(J)^2 \rho(J) \nn \\
&=& 2\, I_2 (2 \sqrt{J} ) K_2 (2 \sqrt{J} ). \ena

As it is well known, the overlap of two coherent states does not vanish in general. Explicitly, we
have \be
\langle J', \gamma '\;| \; J, \gamma \rangle = \frac{2}{N(J)N(J')} 
\sum_{n\geqslant 0} \frac {(J J')^{\frac n2}}{n!(n+2)!}\; e^{-in(n+2)(\gamma - \gamma ')}
\label{6.28}
\en
If $\gamma = \gamma '$, we obtain a Bessel function \be \langle J', \gamma \;|\;J, \gamma \rangle 
= \frac {2} {(J J')^{\frac 12} N(J) N(J')}\; I_2 \; (2(J J')^{\frac 14 } ). \label{Jgamma}
\en

If $\gamma \neq \gamma '$, we can give an integral representation of (\ref{6.28}) in terms of a Theta
function and Bessel functions:\cite{magnus-ob}
\bea
\langle \,J', \gamma' \; | \; J, \gamma \,\rangle &=& \frac{e^{i( \gamma - \gamma ')/4}} {i \pi N(J)\,N(J')}\;
\int_0 ^\pi d\varphi \quad \theta_1 (\frac {\varphi}{\pi} , - \frac{\gamma - \gamma '}{\pi})\;\times \nn\\
&& \makebox[1cm]{} \times \left [
\frac{-e^{-i(\varphi - \gamma + \gamma ')}}{(JJ')^{\frac 12}} \; I_2 \left(2(JJ')^{\frac 14} \; 
e^{i(\varphi - \frac{\gamma - \gamma '}{2} + \frac {\pi}{2})}\right) \right. \nn\\
&&
\makebox[2cm]{} + \left.\frac {e^{i(\varphi + \gamma - \gamma ')}}{(J J')^{\frac 12}} \; I_2 \left(2(JJ')^{\frac 14}\; 
e^{-i(\varphi + \frac {\gamma - \gamma '}{2} - \frac {\pi}{2})}\right)\right ] \ena


\section{COHERENT STATES FOR THE P\"{O}SCHL--TELLER POTENTIALS} 

The relations (\ref{rho})-(\ref{jgamma}) of the previous section are easily generalized to the present case. We
shall list them without unnecessary comments. 

From the energies $E_n=\hbar\, \omega\, e_n$ given by (\ref{hbar}), we get the moments
\begin{equation}
\rho_n = e_1\; e_2\; ...\; e_n = n!\, \frac{\Gamma(n + \nu +1)}{\Gamma (\nu +1)}
\end{equation}
with $\nu = \lambda + \kappa >2$.

Thus, the coherent states read
\begin{equation}
|J,\gamma \rangle = \frac{[\Gamma(\nu+1)]^{1/2}}{N(J)} \sum_{n\geq \,0}\;\; 
\frac{J^{{n}/{2}}e^{-i\gamma n(n+\nu)}}{[n!\, \Gamma(n+\nu +1)]^{\frac12}}\;|n\rangle.
\label{cohstPT}
\end{equation}
The normalization is then given by
\begin{eqnarray}
N(J)^2 &=& \Gamma (\nu +1) \sum_{n \geqslant\,0}\frac{J^n}{n! \, \Gamma (n + \nu +1)}\nonumber\\
&=& \frac{\Gamma (\nu +1)}{J^{{\nu}/{2}}} I_\nu (2 \sqrt{J}). \end{eqnarray} The radius of convergence $R$ is infinite. 
The interval of variation of the evolution parameter $\gamma$
is generically the whole real line, unless the parameter $\nu$ is an integer. 

The numbers $\rho_n$ are moments of a probability distribution $\rho (u)$ involving the modified Bessel function
$K_\nu$ :
\begin{equation}
\rho_n = \int_0^\infty u^n\,\rho (u) du, 
\end{equation}
with (compare with (\ref{rhomeas}))
\be
\rho(u) = \frac{2}{\Gamma (\nu +1)}\; u^{{\nu}/{ 2}}K_\nu (2\sqrt { u}). 
\en
It might be useful to recall here the well-known relation between modified Bessel functions \cite{magnus-ob},
\be
K_\nu (z) = \frac{\pi}{2 \sin\pi\nu} [I_{-\nu} (z) - I_\nu (z)], \; \nu \not\in \Z.
\en
The resolution of the unity is then explicitly given by 
\be
I = \int |J, \gamma \rangle\, \langle J\, , \gamma \, | \, d\mu (J, \gamma),
\en
with
\be
\int (\cdot) d\mu (J, \gamma ) = \lim_{\Gamma \rightarrow \infty} \frac{1}{2 \Gamma}\int_{-\Gamma}^\Gamma \, d\gamma\; 
\left[\; \int_{0}^{+ \infty} k(J)\, (\cdot) \,dJ \; \right],
\en
where
\begin{eqnarray*}
k(J) &=& N(J)^2 \rho (J) \\
&=& 2\, I_{\nu} (2\sqrt J) K_{\nu} (2\sqrt J) . \end{eqnarray*}

Finally, the overlap between two coherent states is given by the series \begin{equation}
\langle J',\gamma ' \, | \, J, \gamma \rangle = \frac{\Gamma (\nu+1)}{N(J)\, N(J')}\;
\sum_{n \geqslant \, 0}\,
\frac{(JJ')^{n/2}}{n! \Gamma (n+ \nu +1)}e^{-in(n+\nu ) (\gamma - \gamma ')}, \end{equation}
which reduces to a Bessel function for $\gamma = \gamma '$: \begin{equation}
\langle J',\gamma \, | \, J, \gamma \rangle = \frac{\Gamma (\nu +1)}{N(J) N(J')}\,
\frac{I_{\nu} (2(JJ')^{1/4})} {(JJ')^{\nu/4}}. \end{equation}

At this point, we should emphasize the fact that, when $\gamma = 0$ and $J$ is taken as a complex parameter, our
temporally stable families of coherent states (\ref{jgamma}) and (\ref{cohstPT}) are nothing else but the temporal
evolution orbits of the well-known Barut--Girardello coherent states for $SU(1,1)$ \cite{barut}.   It seems that this
connection between infinite square-well/P\"oschl--Teller potentials and the latter CS has not been pointed out so far.

In addition, we should also quote Nieto and Simmons \cite{nieto},  who have  considered the infinite square-well
and the P\"oschl--Teller potentials as examples of their construction of coherent states. The latter are required to
minimize an uncertainty relation or, equivalently, to be eigenvectors of some ``lowering operator'' $A^-$ ({\em \`a la}
Barut--Girardello \cite{barut}).  However,  those states have a totally different meaning and should be considered only 
in the semiclassical limit.

\section{PHYSICAL FEATURES OF THE COHERENT STATES} 

In this section, we shall study  the spatial and temporal features of the coherent states, treating 
together     the infinite well  CS (\ref{cohst-well}) and   the  P\"oschl--Teller CS  (\ref{cohstPT}),  
the former being obtained from the latter simply by putting  $\nu = \lambda + \kappa = 2$.
As (infinite) superposition of stationary states which are spatially and temporally periodic  for  integer values of
$\nu $, they should display nonambiguous revivals and fractional revivals. Quantum revivals have recently attracted the
interest of many authors and some of them have considered the infinite square-well as a toy-model for preparing more
realistic studies. But let us first recall the main definitions concerning the notion of revival, as given in  
\cite{aron-stroud}. For other related works, see   \cite{aver,kinzel,matos,gross-rost,stifter,marzo}; for updated references,
see also \cite{robinett}. 

A {\em revival} of a wave function occurs when a wave function evolves in time to a state closely reproducing its
initial form. A {\em fractional revival} occurs when the wave function evolves in time to a state that can be
described as a collection of spatially distributed sub-wave functions, each of which closely reproduces the shape of 
the initial wave function. If a revival corresponds to phase alignments of nearest-neighbor energy eigenstates that
constitute the wave function, it can be asserted that a fractional revival corresponds to phase alignments of
nonadjacent energy eigenstates that constitute this wave function.

For a general wave packet of the form
\be
|\psi (t)\rangle = \sum_{n \geqslant 0}c_n\, e^{-iE_n {t}/{\hbar}} |n\rangle, 
\label{expansion}
\en
with $ \sum_{n \geqslant 0} |c_n|^2 = 1$, the concept of revival arises from the weighting probabilities $|c_n|^2$. 
Suppose that the expansion (\ref{expansion}) is strongly weighted around a mean value $\langle n \rangle$ for the
number operator $N, \, N |n\rangle = n |n\rangle$: \be
\langle \psi | N | \psi \rangle = \sum_{n \geqslant 0} n \,|c_n|^2 \equiv \langle n \rangle.
\label{mean}
\en
Let $\bar n \in \N$ be the closest integer to $\langle n \rangle$. Assuming that the spread
$\sigma \approx \Delta n \equiv\left[ \langle n^2 \rangle - \langle n \rangle ^2 \right]^{1/2}$
is small compared with $\langle n \rangle \approx \bar n$, we expand the energy $E_n$ in a Taylor series in $n$ around 
the centrally excited value $\bar n$:
\be
E_n \simeq E_{\bar n} + E'_{\bar n}(n- \bar n) + \frac 12 E''_{\bar n}(n- \bar n)^2 + \frac 16 E'''_{\bar n}(n- \bar n)^3 + 
\ldots,
\label{energy-exp}
\en
where each prime on $E_{\bar n}$ denotes a derivative. These derivatives define distinct time scales \cite{aver},  
namely the {\em classical period} $T_{\rm cl} = {2\pi \hbar}/{|E'_{\bar n}|}$; the {\em revival time} $t_{\rm rev} = {2\pi
\hbar}/{\frac 12|E''_{\bar n}|}$; the {\em superrevival time}  $t_{\rm sr} = {2\pi \hbar}/{\frac 16|E'''_{\bar n}|}$; and so
on. Inserting this expansion into the the evolution factor $e^{-iE_n {t}/{\hbar}}$ of (\ref{expansion}) allows us to
understand the possible occurrence of a quasiperiodic revival structure of the wave packet (\ref{expansion}) {\em
according to} the weighting probability $n \mapsto |c_n|^2$. In the present case, we have 
\be 
E_n = \frac{\hbar}{2ma^2}\, n(n+\nu) =
\frac{\hbar}{2ma^2}\,\left[{\bar n}({\bar n}+\nu) + (2{\bar n}+\nu)(n- \bar n) + (n- \bar n)^2 \right]. 
\en 
So the first
characteristic time is the  ``classical'' period 
\be 
T_{\rm cl} = \frac{2\pi \hbar}{2{\bar n}+\nu}\; \frac{2ma^2}{\hbar^2} =
\frac{2\pi ma^2}{\hbar(\bar n + \frac{\nu}{2})}, 
\en 
which should be compared with the actual classical (Bohr-Sommerfeld)
counterpart deduced from (\ref{PTperiod}) and (\ref{PTaction}),
\be
T = \frac{2\pi ma^2}{A + a \sqrt{mV_o} [\sqrt{\lambda (\lambda-1)} + \sqrt{\kappa (\kappa -1)}]}.
\en
The second characteristic time is the revival time \be
t_{\rm rev} = \frac{4\pi ma^2}{\hbar} = (2{\bar n}+\nu)T_{\rm cl}. \en
There is no superrevival time here, because the energy is a quadratic function of $n$.

With these  definitions, the wave packet (\ref{expansion}) reads in the present situation (up to a global phase factor):
\be
|\psi (t)\rangle = \sum_{n \geqslant 0}c_n\, 
e^{-2\pi i \left[ (n-{\bar n})\frac{t}{T_{\rm cl}} + (n-{\bar n})^2 \frac{t}{t_{\rm rev}} \right]}  |n\rangle. 
\label{expansion2}
\en
Hence, it will undergo motion with the classical period, modulated by the revival phase \cite{bluhm}.
Since $T_{\rm cl} \ll t_{\rm rev}$ for large $\bar n$, the classical period dominates for small values of $t$ 
(mod $t_{\rm rev}$), and the motion is then periodic with period $T_{\rm cl} $. As $t$ increases from 0 and becomes
nonnegligible with respect to $t_{\rm rev}$, the revival term $(n-{\bar n})^2\frac{t}{t_{\rm rev}}$ in the phase of
(\ref{expansion2}) causes the wave packet to spread and collapse. The latter gathers into a series of subsidiary waves, the
fractional revivals, which move periodically with a period equal to a rational fraction of $T_{\rm cl} $.
Then, a full revival obviously occurs at each multiple of   $t_{\rm rev}$. 

In order to put into evidence these revival structures for a given wave packet 
$\psi (x,t) = \langle x | \psi (t)\rangle$, an efficient method is to calculate its autocorrelation
 function \cite{bluhm} 
\bea
A(t) &=& \langle \psi (x,0) \; | \psi (x,t) \rangle \nn \\ && \nn \\ 
&=& \sum_{n \geqslant 0} |c_n|^2\; e^{-iE_n {t}/{\hbar}} . 
\ena
Numerically, $|A(t) |^2$ varies between 0 and 1. The maximum $|A(t)|^2 = 1$ is reached when $\psi (x,t)$ exactly
matches the initial wave packet $\psi (x,0)$, and the minimum 0 corresponds to nonoverlapping: $\psi (x,t)$ is far
from the initial state. On the other hand, fractional revivals and fractional ``superrevivals" appear (in the general case) 
as periodic peaks  in $|A(t) |^2$ with periods that are rational fractions of the classical round trip time $T_{\rm cl} $
 and the revival time $t_{\rm rev}$. 

Since the weighting distribution $|c_n|^2$ is crucial for understanding the temporal behavior of the wave packet
(\ref{expansion}), it is worthwhile to give also some general precisions of a
 statistical nature \cite{sol,mandel,perina}. 
 before examining the special case of our coherent states. It is clear that the revival features will be more or less
apparent, depending on the value of the deviation $(n- \bar n)$ (relatively to $n$) that is effectively taken into
account in the construction of the wave packet. In this respect, it is interesting to compare $|c_n|^2$ with the
Poissonian case $\langle n \rangle^n \,e^{-\langle n \rangle} / n!$ and with the Gaussian case,
$(2\pi (\Delta n)^2)^{-1/2} \exp[- {(n-\langle n \rangle)^2}/{2(\Delta n)^2}]$.

A quantitative estimate is given by the so-called Mandel parameter $Q$ \cite{sol,mandel} defined as follows:
\be 
Q = \frac{(\Delta n)^2}{\langle n \rangle} - 1. 
\en 
In the Poissonian case, we have $Q = 0$, i.e. $\Delta n =
{\langle n \rangle}^{1/2}$. We say that the weighting distribution  is sub-Poissonian (resp. super-Poissonian) if $Q
< 0$ (resp. $Q > 0$). In the super-Poissonian case, {\em i.e.}, $\Delta n > {\langle n \rangle}^{1/2}$, the set of
states
$| n \rangle$ which contribute significantly to the wave packet can be rather widely spread around $n \simeq \langle n
\rangle$, and this may have important consequences for the properties of localization and temporal stability of the
wave packet.

When the wave packets are precisely our coherent states 
\be
|J,\gamma \rangle = \frac{1}{N(J)} \sum_{n\geqslant \,0}\; \frac{J^{n/2}\, 
e^{-i e_n \gamma}}{\sqrt{\rho_n}}\;|n\rangle, 
\label{cohstPT2}
\en
the weighting distribution depends on $J$, 
\be
|c_n|^2 = \frac{J^{n}}{N(J)^2 \,\rho_n },
 \en
and we can see the interesting statistical interplay with the probability distribution $\rho(J)$ of which the
$\rho_n$ are the moments, see (\ref{rho_n}). 

\begin{figure}
\centering
  \begin{center}
\vspace*{-2cm}
\includegraphics[width=8cm,origin=c,angle=-90]{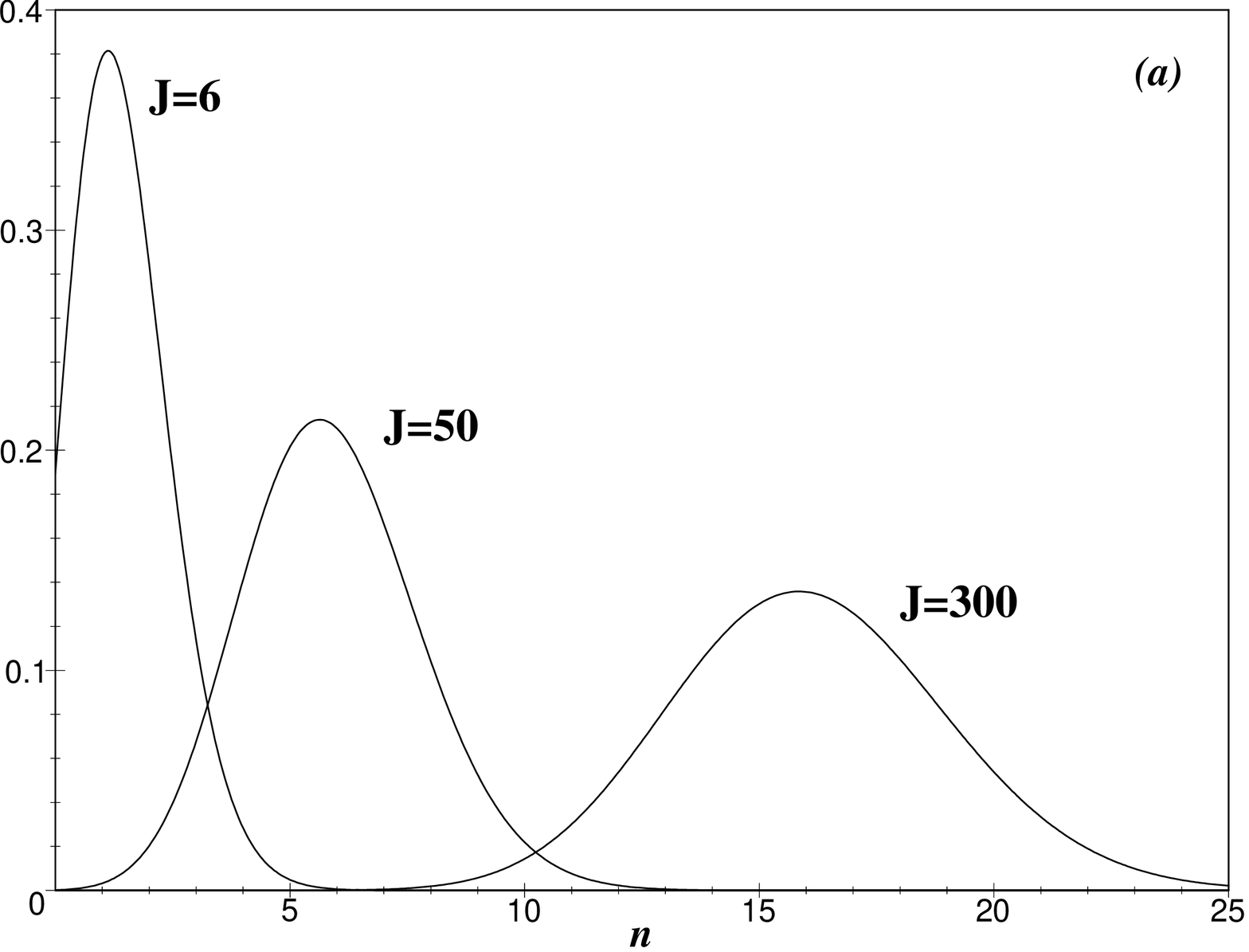}  
\\
\vspace*{-1cm} 
\includegraphics[width=8cm,origin=c,angle=270]{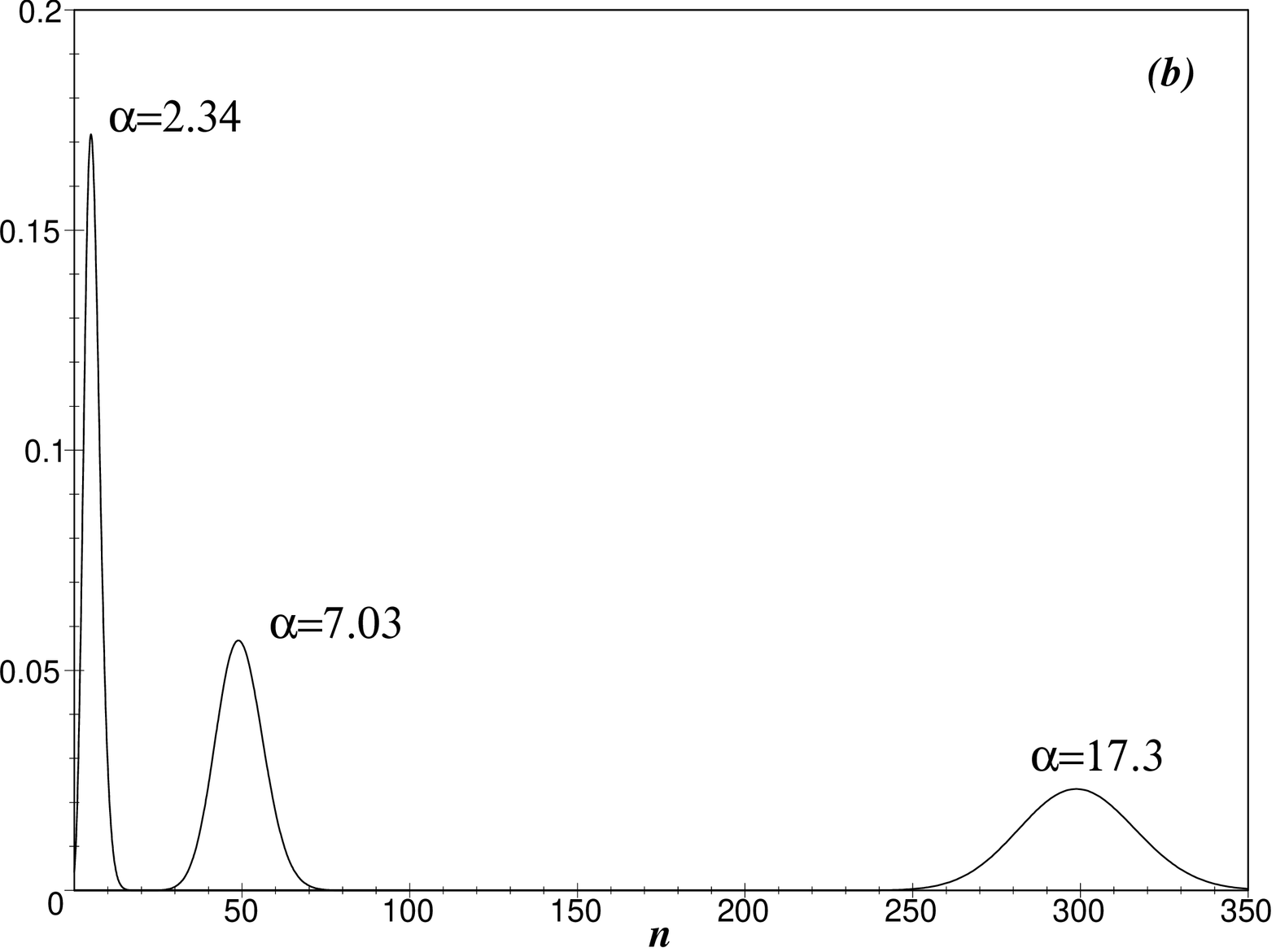} 
\vspace*{-1cm} 
  \end{center}
\caption{(a) The weighting distribution  $|c_n |^2 \equiv  D(n,J, \nu )   $ given in (\ref{distrPT}) for 
the infinite square-well  $\nu = 2$ and different
values of $J$. Note the almost Gaussian shape at $J=300$,  centered at 
$n = \langle n \rangle =\sqrt{J} -\frac\nu 2 - \frac 14 \simeq 16$, a  width equal to 
$2 \Delta n =   \sqrt{2}  J^{1/4}\simeq  5.9$;
 (b) The same for for the harmonic oscillator: $|c_n |^2 = \frac {1}{n!} |\alpha|^{2n} e^{-|\alpha|^{2}}.$ 
The values of $\alpha$ are chosen so 
as to get essentially the same mean energy values as in (a): $\alpha = \sqrt{J}$.}
\label{figure11}
\end{figure}

To simplify, we put
\bea
E(J) &\equiv& N(J)^2 \;\;=\;\; \sum_{n\geqslant \,0}\;\frac{J^{n}}{\rho_n} \nn \\
&=& \frac{\Gamma(\nu + 1)}{J^{\nu/2}} \; I_{\nu}(2 \sqrt{J}), \; \nu \geqslant 2.
\ena
The following mean values are easily computed, together with their asymptotic values for large $J$ \cite{magnus-ob}, 
\bea
{\langle n \rangle} &=& \frac{J}{E(J)} \frac{d}{dJ}E(J) \;\;=\;\; J \frac{d}{dJ} \ln E(J) \nn \\
&=& \sqrt{J} \frac{I_{\nu+1}(2 \sqrt{J})}{I_{\nu}(2 \sqrt{J})} 
= \sqrt{J} -\frac\nu 2 - \frac 14 + {O} (\frac{1}{ \sqrt{J}}). 
\\ && \makebox[1cm]{} \nn \\
&& \makebox[1cm]{} \nn \\
{\langle n^2 \rangle} &=& \frac{J}{E(J)} \frac{d}{dJ} J \frac{d}{dJ} E(J) \nn \\
&=& \sqrt{J} \frac{I_{\nu+1}(2 \sqrt{J})}{I_{\nu}(2 \sqrt{J}) } + 
J \frac{I_{\nu+2}(2 \sqrt{J})}{I_{\nu}(2 \sqrt{J})} \nn \\ &=& {\langle n \rangle} + 
J \frac{I_{\nu+2}(2 \sqrt{J})}{I_{\nu}(2 \sqrt{J})}
\approx \sqrt{J} (\sqrt{J} +1) \quad (J\gg 1).
\ena
So, the dispersion is
\bea
 (\Delta n)^2 &=& J \; \frac{I_{\nu+2}(2 \sqrt{J})}{I_{\nu}(2 \sqrt{J})} + 
       {\langle n \rangle} - {\langle n \rangle}^2  \nn
\\ &&  \nn
\\ &=& \frac{J}{[I_{\nu}(2 \sqrt{J})]^2} \,
 \left( I_{\nu+2}(2 \sqrt{J}) \,I_{\nu}(2 \sqrt{J}) - [I_{\nu+1}(2 \sqrt{J})]^2 \right)
+ \sqrt{J}\;\frac{I_{\nu+1}(2 \sqrt{J})}{I_{\nu}(2 \sqrt{J}) } \nn
\\ &&  \nn
\\ &\approx& 
 \frac{\sqrt{J}}{2}, \quad \mbox{for $J$ large.}
\ena
Finally, the Mandel parameter is given explicitly by 
\bea
Q &=& J \frac{d}{dJ} \ln \frac{d}{dJ} \ln E(J) \nn \\ && \makebox[1cm]{} \nn \\ 
&=& \sqrt{J}\;\left[ \frac{I_{\nu+2}(2 \sqrt{J})}{I_{\nu+1}(2 \sqrt{J})}
- \frac{I_{\nu+1}(2 \sqrt{J})}{I_{\nu}(2 \sqrt{J})}\right]. 
\ena
It is easily checked that 
$(I_{\nu+1}(x))^2 \geqslant I_{\nu}(x) I_{\nu+2}(x)$, for any $x\geqslant 0$, and thus,
 $Q \leqslant 0$ for any $J \geqslant 0$.  Note that $Q \simeq 0$ for large $J$,
 while $Q \simeq -J$
 for small $J$. Therefore, $|c_n|^2 $ is sub-Poissonian in the case of our coherent states, whereas a
quasi-Poissonian behavior is restored at high $J$. This fact is important for understanding the curves presented in
Figure \ref{figure11}(a), which show  the distributions 
\be
D(n,J,\nu) \equiv |c_n|^2  =
\frac{1}{n! \, \Gamma(n+\nu+1)}\frac{J^{n+\nu/2}}{I_\nu(2\sqrt{J})} 
\label{distrPT}
\en
 for $\nu = 2$  and different
values of $J$. For the sake of comparison, we show in Figure \ref{figure11}(b) 
the corresponding distribution $|c_n |^2 = \frac {1}{n!} |\alpha|^{2n} e^{-|\alpha|^{2}}$ for the harmonic oscillator.
Exactly as in the latter case, it can be shown easily that the  distribution $D(n,J,\nu)$ tends for $J \to \infty$ to a
Gaussian distribution. This Gaussian is centered at $\sqrt{J} -\frac\nu 2 - \frac 14 $ and has a half-width equal to 
$\frac {1} {\sqrt{2}} J^{1/4} $:
\be
D(n,J,\nu) \approx \frac{1}{\sqrt{\pi \sqrt{J}}}\, e^{-\left[n - (\sqrt{J} -\frac\nu 2 - \frac 14)\right]^2/\sqrt{J}}
\quad (n \gg 1).
\en

\begin{figure}
  \begin{center}
\vspace*{-2cm}
\includegraphics[width=6cm,origin=c,angle=-90]{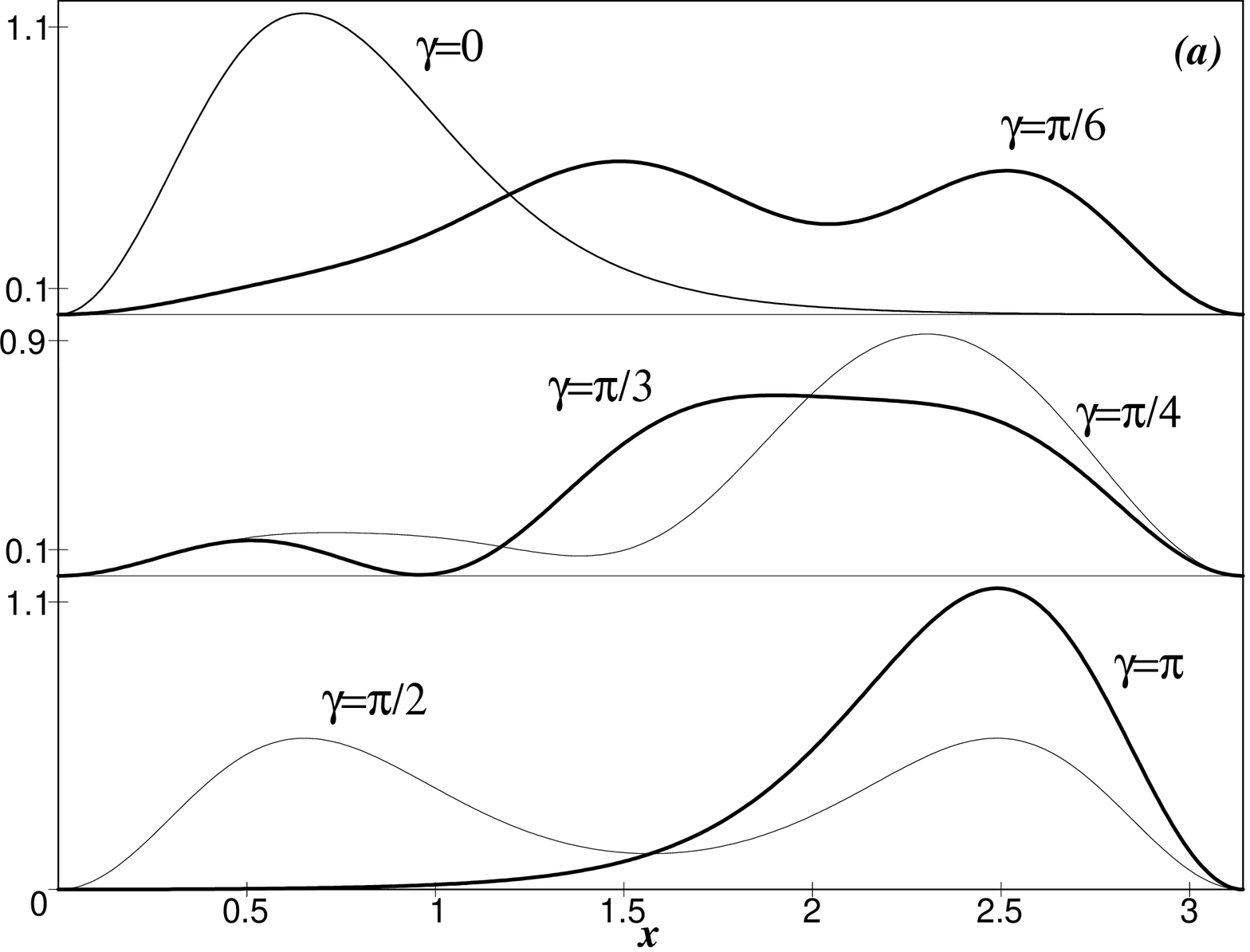}  
\\
\vspace*{-5mm} 
\includegraphics[width=6cm,origin=c,angle=270]{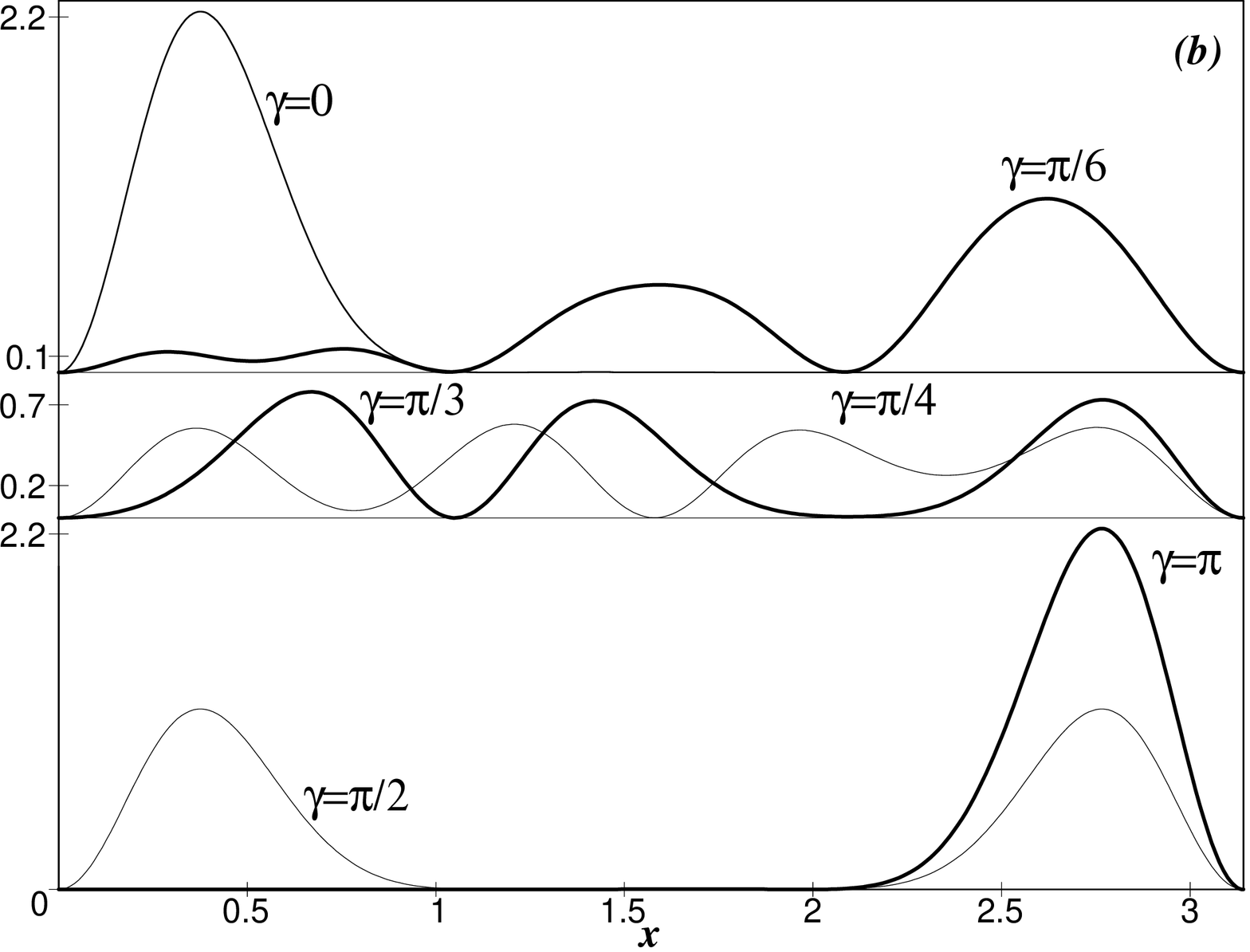} 
\\
\vspace*{-5mm} 
\includegraphics[width=6cm,origin=c,angle=270]{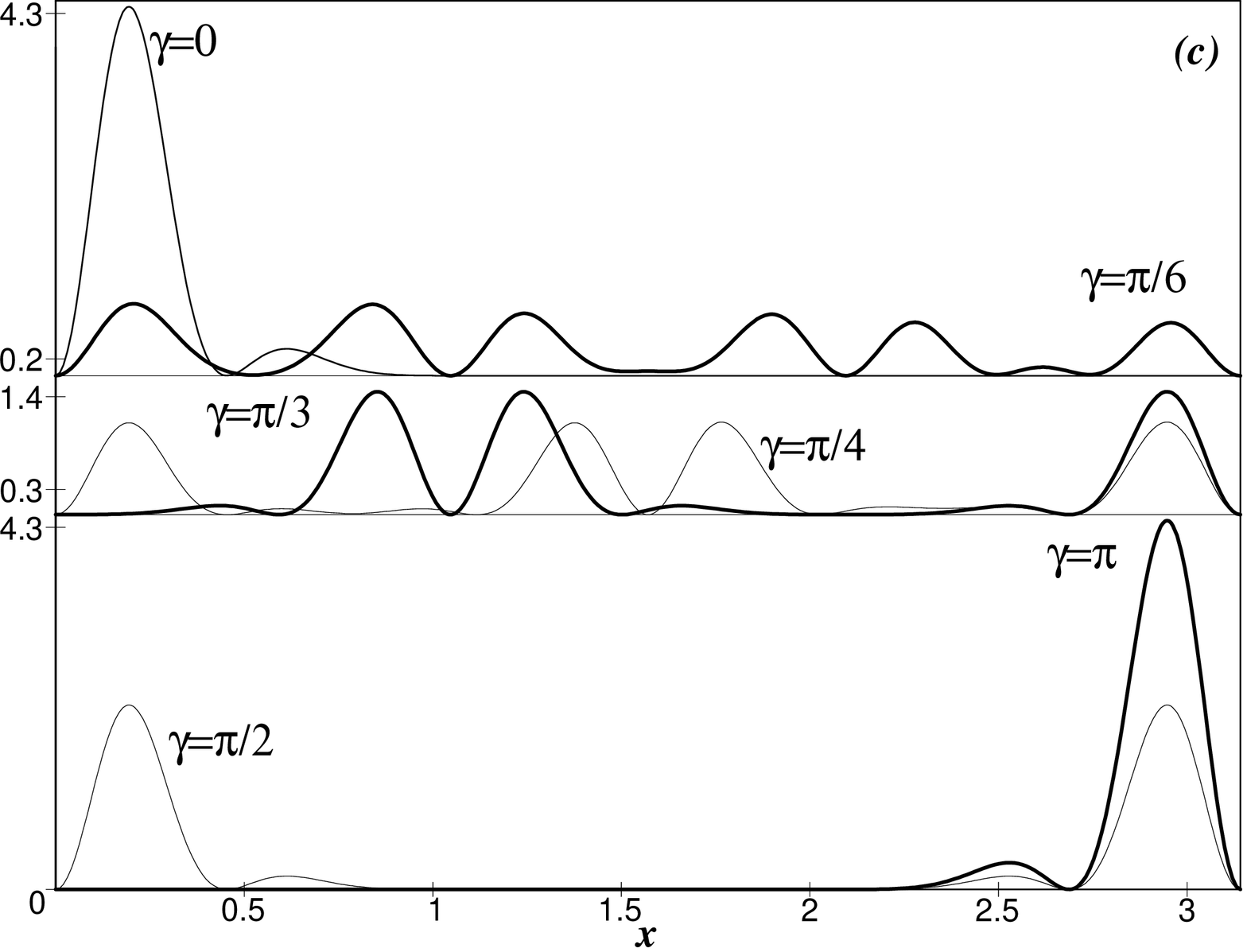} 
\vspace*{-1cm} 
  \end{center}
\caption{The evolution (vs. $\gamma$)
of the probability density $|\langle x | J,\gamma \rangle|^2$, in the case of the infinite square-well
for (a) $J= 2$; (b) $J= 10$; and (c) $J= 50$. 
We note the perfect revival at $\gamma = \pi = \frac 12 t_{\rm rev}$ (in suitable units), 
symmetrically  with respect to the center of the well.} 
\label{figure12}
\end{figure}

We consider now the probability density $|\langle x | J,\gamma \rangle|^2$ as a function 
of the evolution  parameter $\gamma = \omega t$ for increasing values of $J$.  
This evolution is shown in Figure \ref{figure12}  in the case of the infinite square-well, for $J= 2, 10,$ and 50.
We can see here at $\gamma = \pi = \frac 12 t_{\rm rev}$ a perfect  revival  of the initial shape at $\gamma =0$.
This revival takes place near the opposite wall, as expected from the symmetry  with respect to the center of the well.
On the other hand, the ruling of the wave packet evolution by the classical period
$T_{\rm cl} = \dis \frac{t_{\rm rev}}{2 \bar n + \nu} = \frac{\pi}{  \bar n +1}$ becomes more and more apparent as $J$
increases. We also note that, at multiples of the half reversal time $\frac 12 t_{\rm rev} = \pi$, the probability of
localization near the walls increases with the energy $J$.

\begin{figure}
\centering
  \begin{center}
\vspace*{-2cm}
\includegraphics[width=8cm,origin=c,angle=-90]{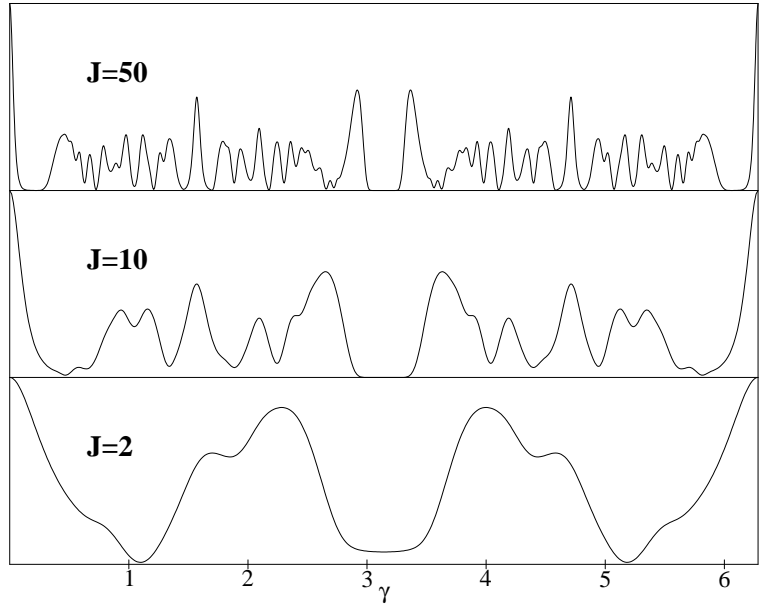}  
  \end{center}
\vspace*{-1cm} 
\caption{Squared modulus $|\langle J,0\, |\, J, \omega t  \rangle |^2$ of the
autocorrelation vs. $\gamma = \omega t$ for the infinite square-well, for $J = 2,10,50$.
As in  Figure \ref{figure12},  the large $J$ regime is characterized by the  occurrence of  fractional revivals.}
\label{figure13}
\end{figure}

\begin{figure}
  \begin{center}
\includegraphics[width=6cm,height=6cm]{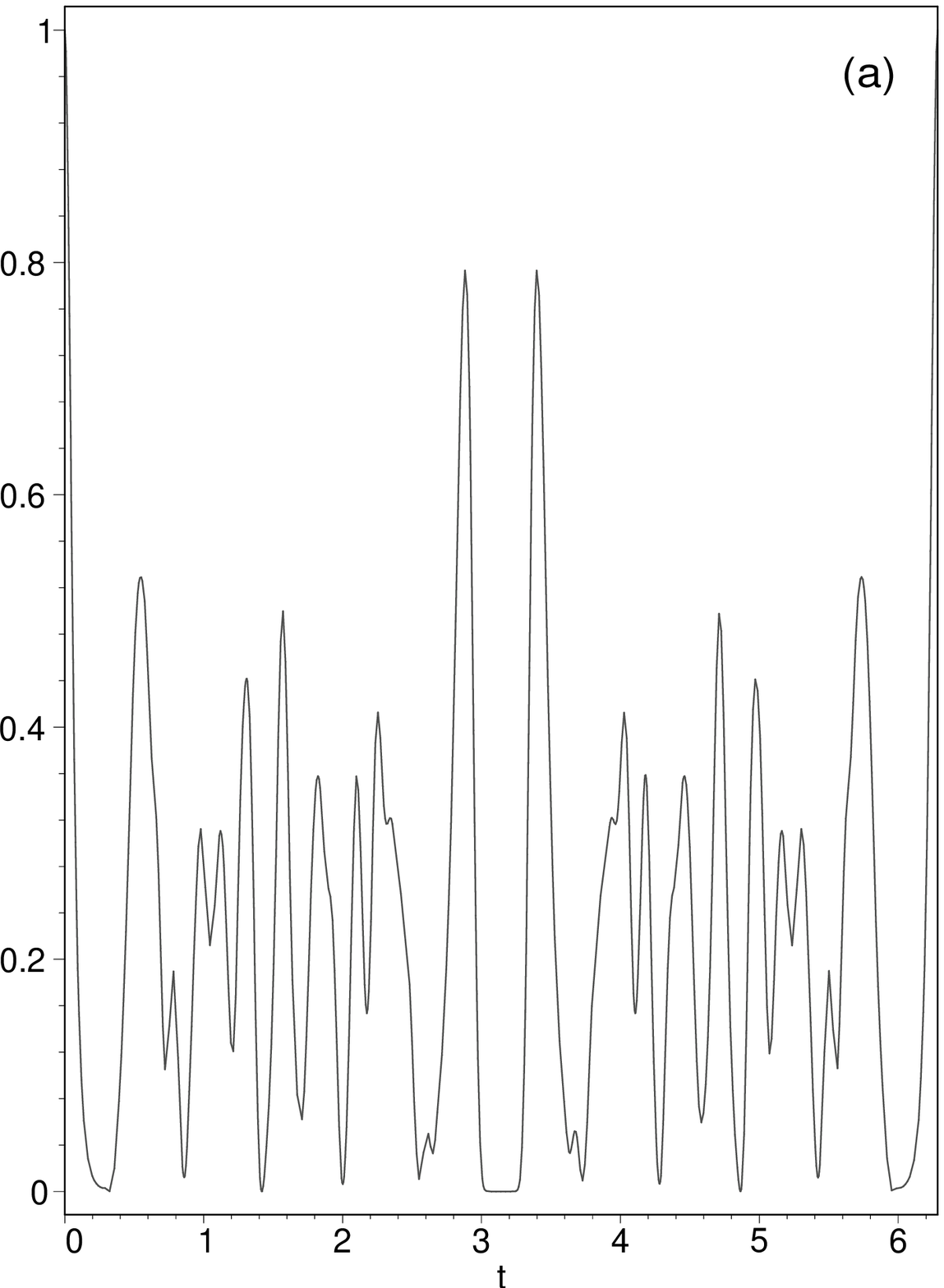}
\hspace{3mm}
\includegraphics[width=6cm,height=6cm]{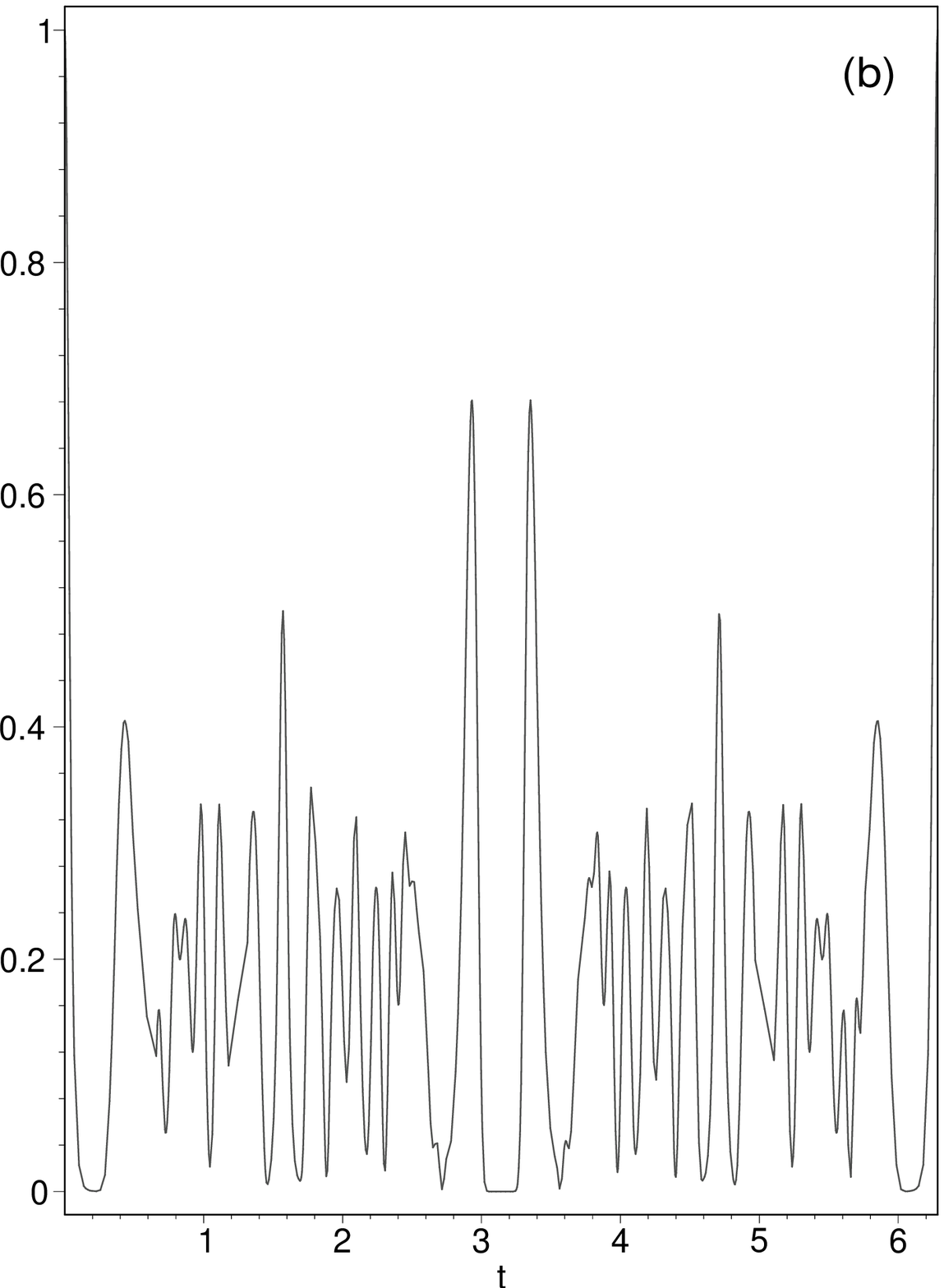}
  \end{center}
\caption{Squared modulus $|\langle J,0\, |\, J, \omega t  \rangle |^2$ of the
autocorrelation   for the P\"{o}schl--Teller potential with  $n_{\rm max=10}$, for (a)  $J = 20$; (b) $J =  40$.}
\label{figure14}
\end{figure}

In Figure \ref{figure13} ,  we show the squared modulus 
\begin{eqnarray}
|\langle J,0|e^{-\frac{i}{\hbar} Ht}\, |\,J,0 \rangle |^2 &=& |\langle J,0\, |\, J, \omega t \rangle |^2\\
&=& \frac{\Gamma(\nu +1)}{N(J)^2}\; \left|\; 
\sum_{n \geqslant \, 0}\, \frac{J^n}{n!\Gamma(n+\nu +1)} e^{-i\omega n(n+\nu)t}\; \right |^2 
\end{eqnarray}
of the autocorrelation vs. $\gamma = \omega t$ for the infinite well, for   $J = 2, 10, 50$. 
Like in  Figure \ref{figure12}, we draw the attention on the large $J$ regime. Here fractional revivals   occur
as intermediate peaks at rational multiples of the classical period
$T_{\rm cl} =  \dis\frac{\pi}{\bar n +1} \approx \dis  \frac{\pi}{\sqrt{J}}, J \gg 1$, and they 
tend  to diminish as $J$ increases, which is clearly the mark of a quasiclassical behavior. 
The same quantity is shown in  Figure \ref{figure14} for the P\"oschl--Teller potential, for $J = 20$ and 40.
Note that, in actual calculations like this, one has to choose a finite number of orthonormal eigenstates of the 
P\"oschl--Teller potential, denoted here by $n_{\rm max}$. Correspondingly, the normalization 
of the coherent state $|J, \gamma \rangle$ has then to be modified as
 \bea
\sum_{p=0}^{n_{\rm max}} \frac{J^p}{p! \,\Gamma(p + \kappa + \lambda)} &=&
\frac{I_{\kappa + \lambda -1}(2\sqrt{J})}{J^{\frac 12 (\kappa + \lambda -1)}}  
 - \frac{J^{n_{\rm max} +1}\;_1\!\!\:F_2(1; n_{\rm max} +2, n_{\rm max} +1  + \kappa + \lambda ; J)}{(n_{\rm max} +1)!
\,\Gamma(n_{\rm max} +1  + \kappa + \lambda )}  \nn \\ &&  
\ena
where $_1\!\!\:F_2$ is the hypergeometric function. 

\begin{figure}
  \begin{center}
\includegraphics[width=7cm,origin=c,angle=-90]{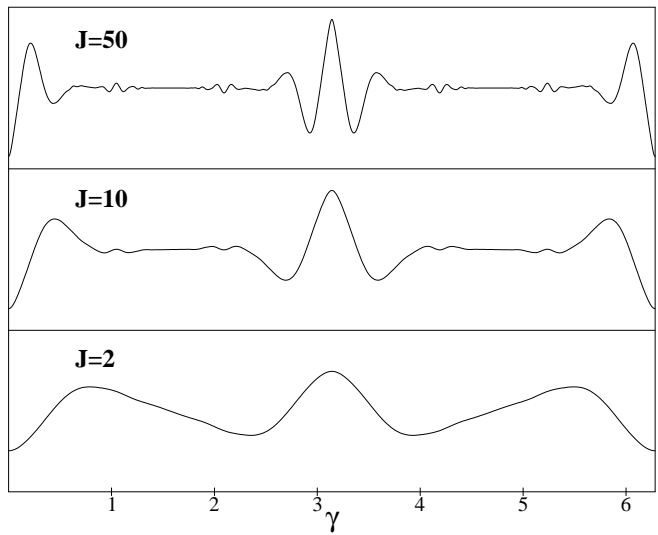}  
  \end{center}
\vspace*{-1.5cm}
\caption{Temporal behavior of the average position of the particle in
the infinite square-well (in the Heisenberg picture), 
$\langle J,0\,|\,Q(t)\,|\,J,0\rangle = \langle J, \omega t = \gamma \,|\,Q\,|\,J, \omega t= \gamma \rangle$,
as a function of $\gamma = \omega t $, for $J = 2,10,50$.}
\label{figure15}
\end{figure}

\begin{figure}
  \begin{center}
\includegraphics[width=6cm,height=6cm]{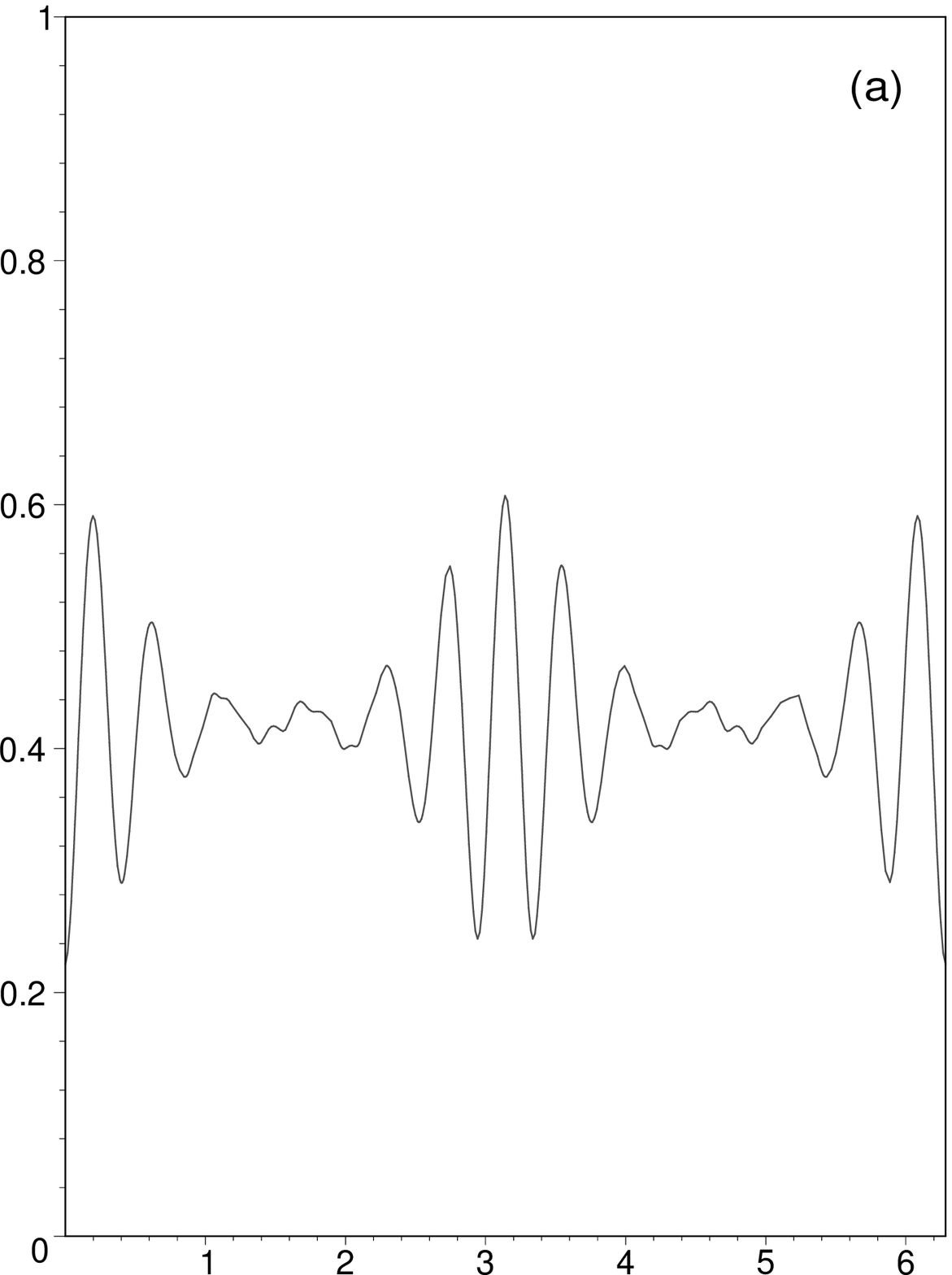}
\hspace{3mm}
\includegraphics[width=6cm,height=6cm]{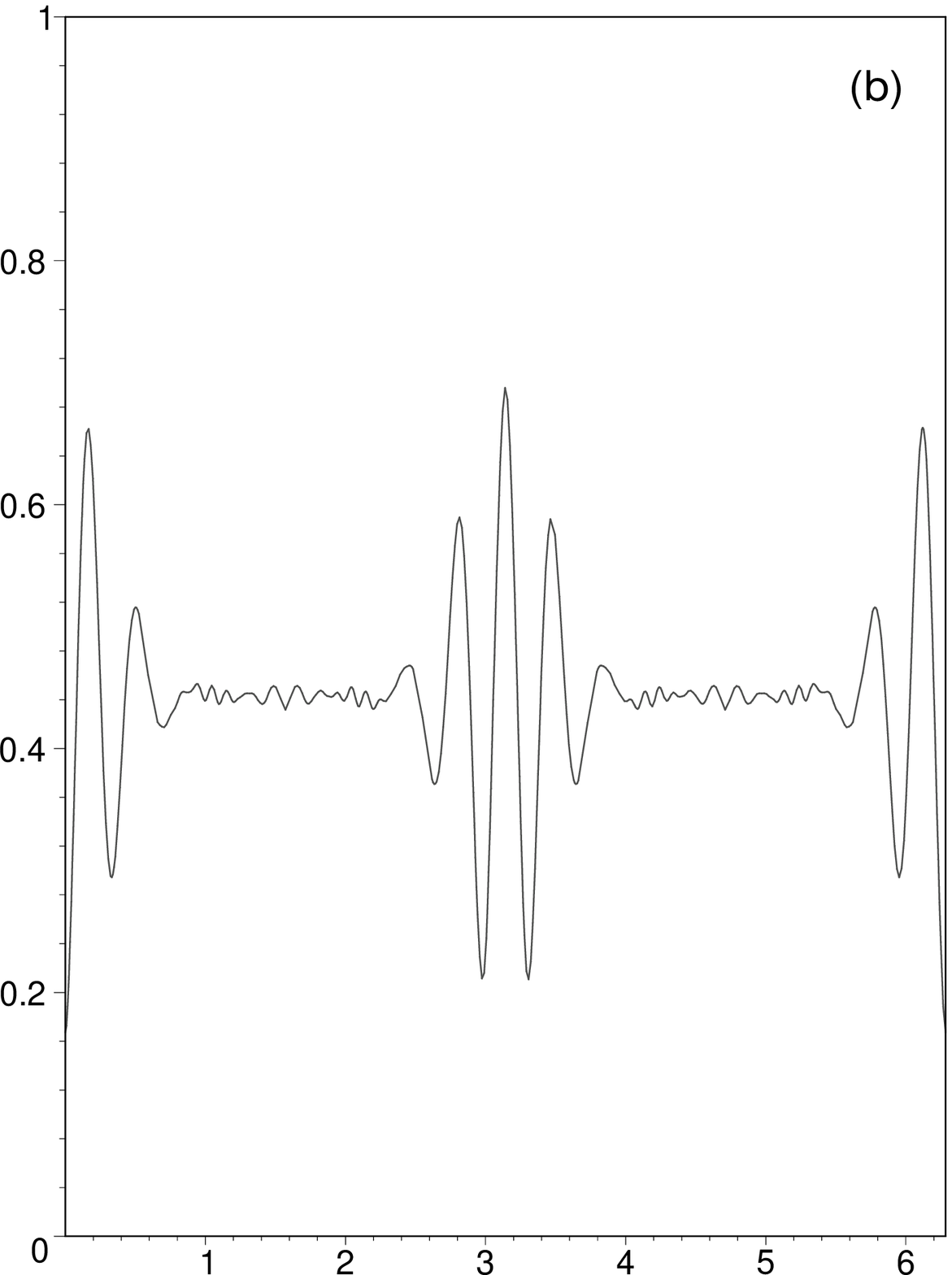}
  \end{center}
\caption{Temporal behavior of the average position for the  asymmetric P\"{o}schl--Teller
potential $(\lambda,\kappa) =   (4,8) $ with  $n_{\rm max=10}$, for (a) $J=20$; and (b) $J =  50$.} 
\label{figure16}
\end{figure}

Most interesting is the temporal behavior of the average position $\langle Q \rangle$ and of
the average momentum $\langle P \rangle$ in such coherent states 
\begin{eqnarray}
\langle J,0\, | \, A(t) \, | \, J,0\rangle &=& \langle J,0 \, | \, e^{\frac{i}{\hbar} Ht}
A e^{-\frac{i}{\hbar} Ht}\, |\, J, 0\rangle \nonumber\\ 
&=& \langle J, \omega t = \gamma \, | \, A \, | \, J, \omega t= \gamma \rangle.
\end{eqnarray}
for $A = Q$ or $A = P \equiv P_1$. This temporal behavior is shown in Figure \ref{figure15} 
for the average position in the infinite square-well, for   $J = 2, 10, 50$.  We note the tendency to
stability around the classical mean value $\frac 12 \pi a$, except for strong oscillations of ultrashort duration between the
walls near $\gamma = n\pi$. The latter increase with $J$ as expected when one approaches the classical regime.
For the sake of comparison, we show in Figure \ref{figure16}  the temporal behavior of the average position in the asymmetric
P\"oschl--Teller potential $(\kappa, \lambda) = (4,8)$, for  $J = 20$ and  50.


\begin{figure}
  \begin{center}
\includegraphics[width=7cm,origin=c,angle=-90]{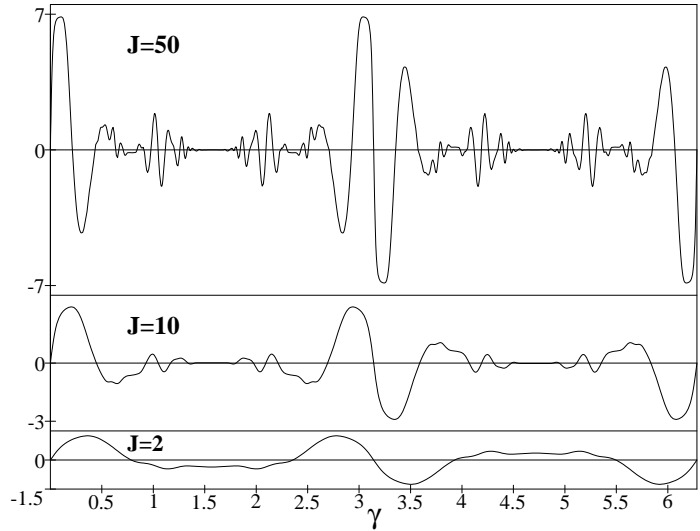}  
  \end{center}
\vspace*{-1cm}
\caption{Temporal behavior of the average momentum $\langle J,0\,| \,P(t) \,|\, J,0\rangle$ in the
case of the infinite square-well, for   $J = 2, 10, 50$.} 
\label{figure17}
\end{figure}

 Figure \ref{figure17} shows the temporal behavior of the average momentum  
$\langle J,0\,| \,P(t) \,|\, J,0\rangle$ in the case of the infinite square-well, for   $J = 2, 10, 50$.
Like in Figure \ref{figure15}, we note the presence of strong ultrashort oscillations at $\gamma = n \pi$,
whereas a tendency to perfect stability (around the classical mean value 0) exists at intermediate values of  $\gamma $
(this tendency is, however, less marked than for the average position).

\begin{figure}
  \begin{center}
\includegraphics[width=7cm,origin=c,angle=-90]{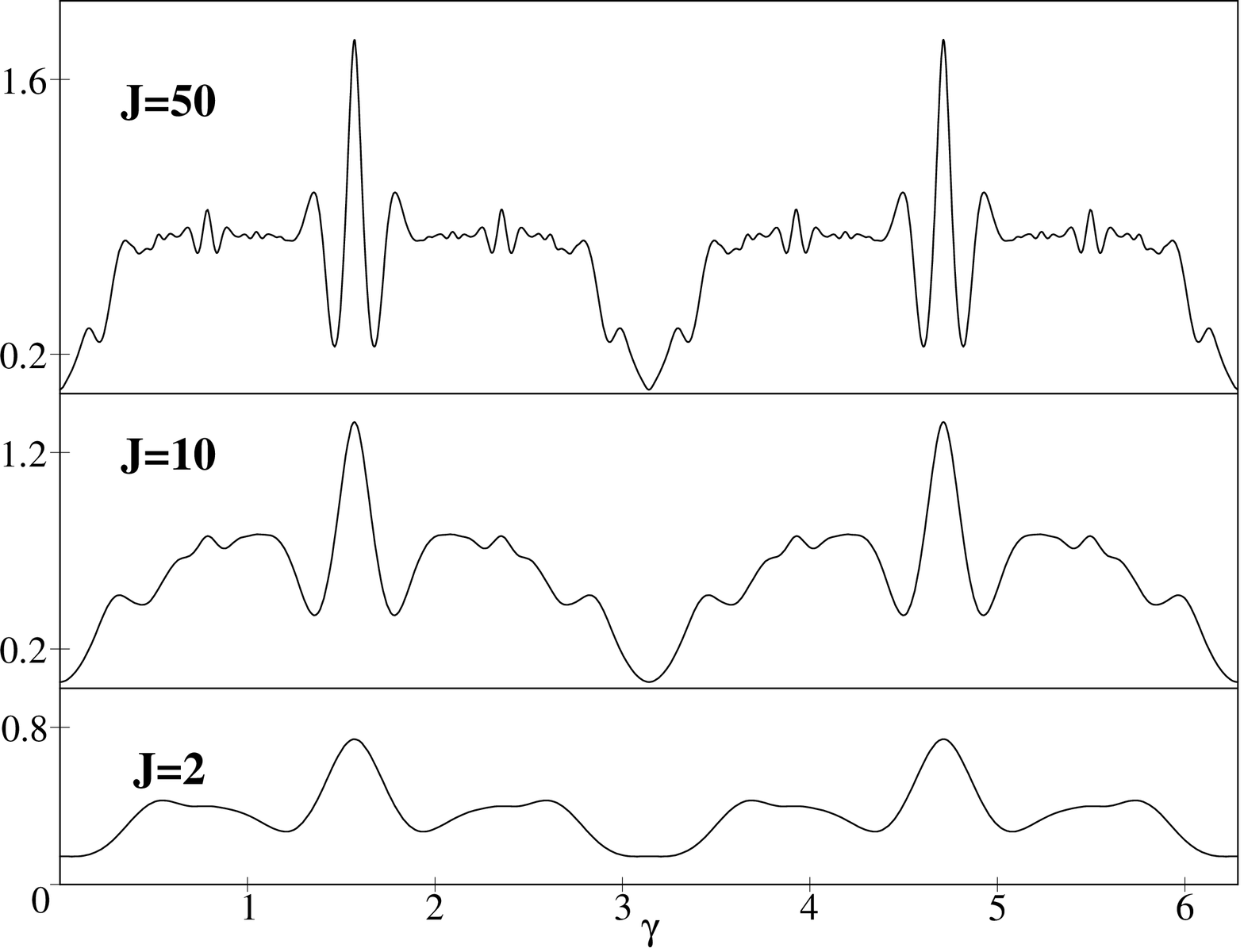}  
  \end{center}
\vspace*{-1cm}
\caption{Temporal behavior of the squared uncertainty in position $(\Delta Q)^2$,
in the case of the infinite square-well, for   $J = 2, 10, 50$.}
\label{figure18}
\end{figure}
\begin{figure}
  \begin{center}
\includegraphics[width=7cm,origin=c,angle=-90]{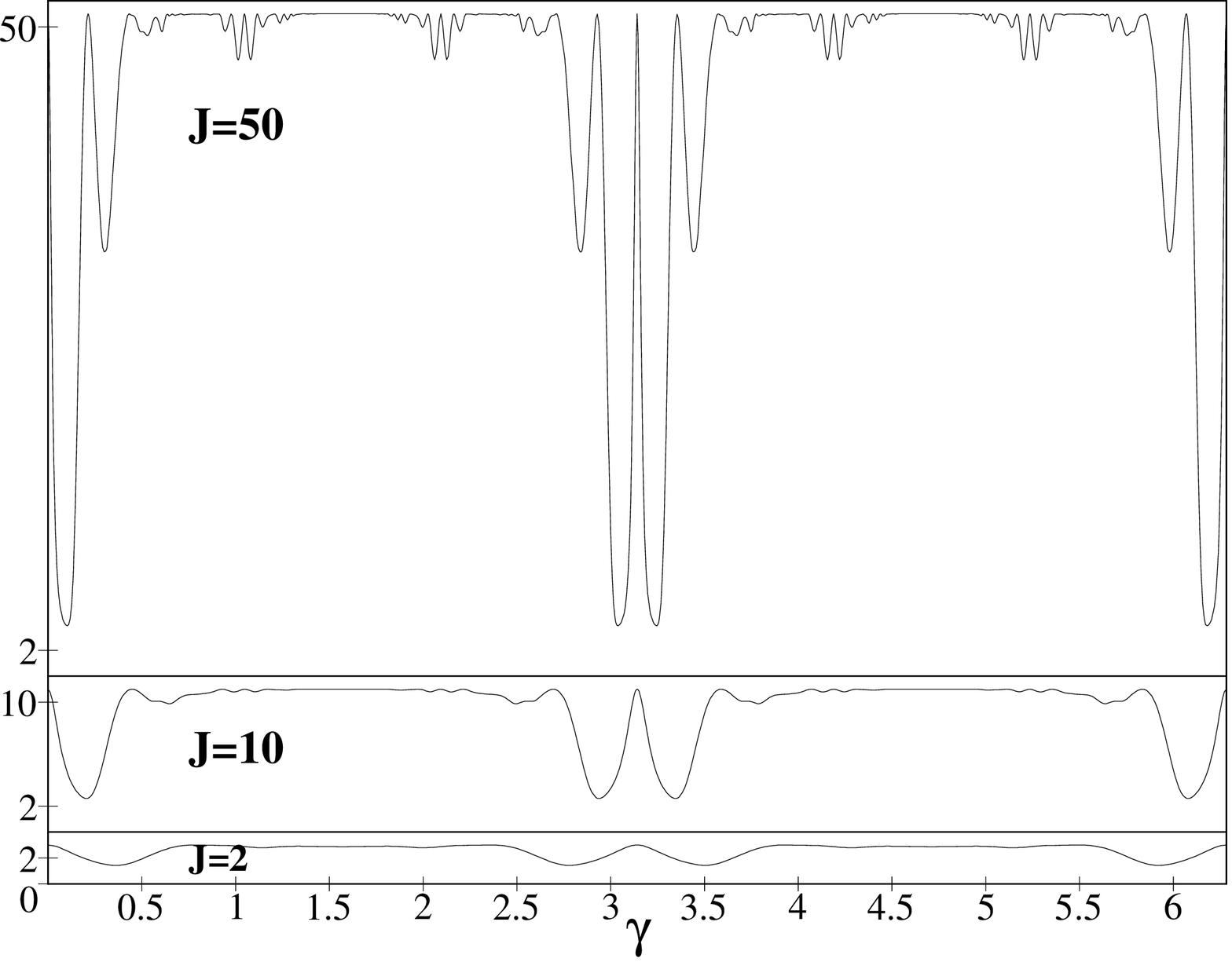}  
  \end{center}
\vspace*{-1cm}
\caption{Temporal behavior of the  squared uncertainty in momentum $(\Delta P)^2$, 
in the case of the infinite square-well, for  $J = 2, 10, 50$.}
\label{figure19}
\end{figure}

\begin{figure}
  \begin{center}
\includegraphics[width=7cm,origin=c,angle=-90]{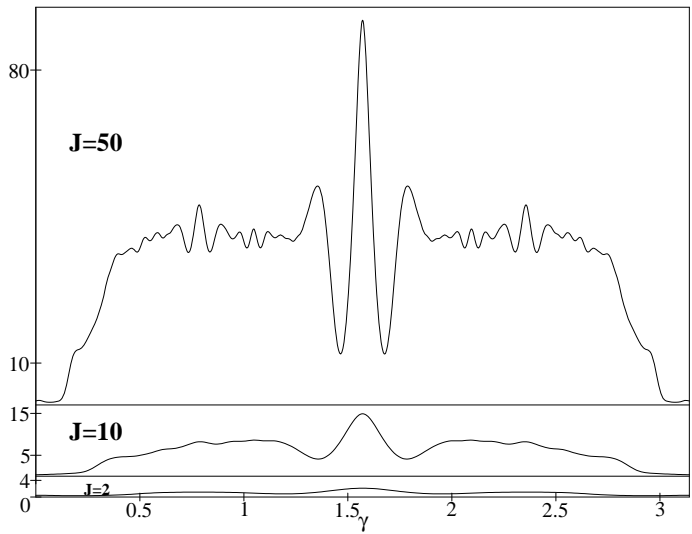}  
  \end{center}
\vspace*{-1cm}
\caption{Temporal behavior of the product of the squared uncertainties $(\Delta Q)^2\,(\Delta P)^2$, 
in the case of the infinite square-well, for  $J = 2, 10, 50$.} 
\label{figure20}
\end{figure}

Next, we examine the  uncertainty in position and momentum, in order to evaluate how close  these CS come 
to saturating the uncertainty relations.   Figure  \ref{figure18} 
  shows the temporal behavior of the  squared uncertainty in position, $(\Delta Q)^2$, for the infinite square-well,
again for   $J = 2, 10, 50$.    Figure  \ref{figure19} 
does the same for momentum,  $(\Delta P)^2$, and
 Figure \ref{figure20} 
shows the product of the two,  $(\Delta Q)^2 (\Delta P)^2$. We note   here that the product approaches the limit value
$\frac 14$ (saturation of the Heisenberg inequality) for a longer time at small $J$. This is consistent with
(\ref{uncertrel}), since at small $J$ the wave packet is centered near the ground state, for which we reach the minimal
value $(0.57)^2$. On the other hand, we also note the strong oscillations of 
$(\Delta Q)^2 (\Delta P)^2$ at half the revival time, a fact which is consistent with the previous figures, showing the
average position and momentum. At $\frac 12 t_{\rm rev}$, the quantum interferences are dominant and they enforce the
spreading   of the  wave packet for a relatively long duration.

\begin{figure} 
  \begin{center}
\includegraphics[width=12cm]{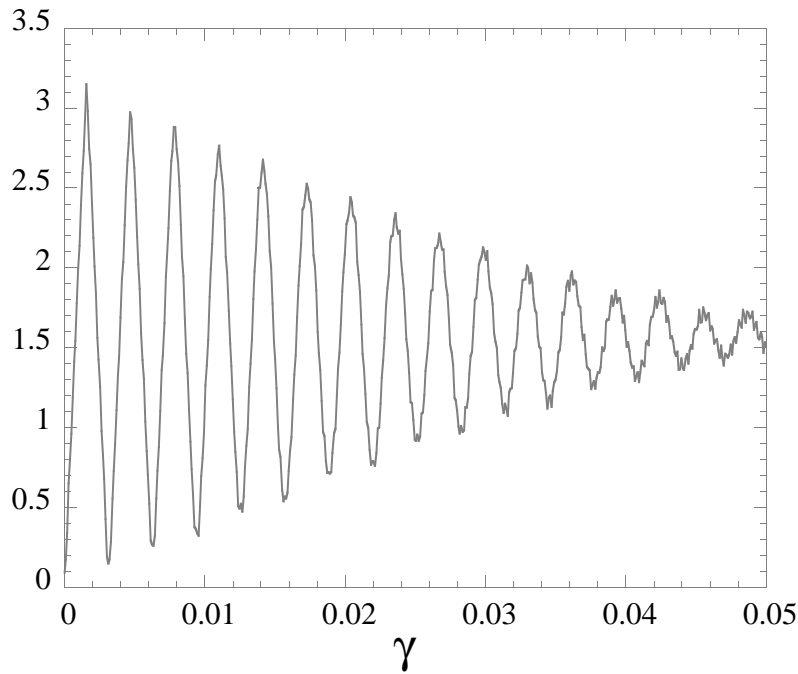}  
  \end{center}
\vspace*{-0.5cm}
\caption{Temporal behavior of the average position in the case of the infinite square-well, for a very high value  $J=
10^6$.} 
\label{figure21}
\end{figure}

As a last information (but not the least!), we exhibit in  Figure  \ref{figure21} 
the temporal behavior of the average position
$\langle J,0\,| \,Q(t) \,|\, J,0\rangle$ for  the infinite square-well, for a very high value $J = 10^6$, near 
$\gamma = \omega t = 0$. Here the quasiclassical behavior is  striking in  the  range of values considered
for $\gamma $. These temporal oscillations are clearly governed by  $T_{\rm cl} \simeq \frac{\pi}{\sqrt{J}} = 3 \times
10^{-3}$ and should be compared with their purely classical counterpart of Figure \ref{figure3}.


\section{DISCUSSION}

Coherent states have many roles to play in quantum theory. Among those roles is included the Hilbert space
representation that coherent states induce, which is largely kinematical in nature, and the adaption of the coherent
states  themselves to some particular dynamics and the possible description that ensues. To accomodate these goals,
the definition of what constitutes a ``set of coherent states'' has been increasingly broadened over the years.
Widening the scope of coherent states also widens the range of potential applications. This basic principle lies
behind the developments in this paper.

The minimal definition of a set of coherent states involves continuity of labeling and a resolution of unity, and, 
therefore, holomorphic representations and/or definitions via groups are just a small subset of the possibilities. In 
the present article, we have exploited this diversity in coherent state definition to study the motion of a particle in
P\"oschl--Teller potentials as well as in the closely related infinite square-well potential.

The specific choices we have made for the set of coherent states are based on two additional guiding principles
besides continuity and resolution of unity \cite{klau2,ga-kl}.  The first of these is ``temporal stability,'' which in 
words asserts that the temporal evolution of any coherent state always remains a coherent state. The second of these,
referred to as the ``action identity'' in   \cite{ga-kl}, chooses variables for the coherent state labels that
have as close a connection as possible with classical ``action-angle'' variables. In particular, for a single degree
of freedom, the label pair
$(J,\gamma)$ is used to identify the coherent state $|J,\gamma\rangle$. Temporal stability means that, under the chosen
dynamics, temporal evolution proceeds according to $|J,\gamma+\omega t\rangle$, for some fixed parameter $\omega$. To
ensure that $(J,\gamma)$ describes action-angle variables, it is sufficient to require that the symplectic potential
induced by the coherent states themselves is of Darboux form, or specifically that
$$
i\hbar\langle J,\gamma|\,d\,|J,\gamma\rangle=J\,d\gamma 
$$
where $d\,|J,\gamma\rangle\equiv|J+dJ,\gamma+d\gamma\rangle-|J, \gamma\rangle$. Temporal stability is what fixes the
{\it phase} behavior of the coherent states,  {\em i.e.}, the factor
$e^{-i\gamma e_n}$ [cf.~(7.15)], while ensuring that $(J,\gamma)$ are canonical  action-angle variables is what fixes
the  {\it amplitude} behavior of the coherent states,  {\em i.e.}, $1/N\sqrt{[e_n]!}$ [cf.~(7.19) and (7.20)]. The
given amplitude behavior may be arrived at by other means \cite{sol},  but requiring that $J$ and $\gamma$ be
canonical classical coordinates is equivalent and tends to stress the physics of the situation.

In order for coherent states to interpolate well between quantum and classical mechanics, it is necessary for values
of the  action $J\gg\hbar$ that the quantum motion be well approximated by the classical motion. In particular, for a
classical system with closed, localized trajectories, a suitable wave packet should, if possible, remain ``coherent''
for a number of classical periods. For the systems under study in this paper, we have demonstrated the tendency for
improved packet coherence with increasing $J$ values within the range studied. For significantly larger values of
$J$, we notice that the packet coherence   substantially improves. Interesting results have been obtained independently
in a related study by Fox and Choi \cite{fox}, who found a similar packet coherence for $10$ or more classical
periods for an infinite square-well, even though they used a different amplitude prescription for their coherent
states. In both works, however, the probability distribution shows a Gaussian behavior for large values of
$J$, and this  explains the similarity of the results.

It would appear that allowing for generalized phase and amplitude behavior in the definition of coherent states has
led us closer to the idealized goal of a set of coherent states  adapted to a chosen system and   having a large
number   of properties in common with the associated classical system, despite being fully quantum in their
characteristics. 
\bigskip

\noi
{\bf Note added:}  After completion of the present paper, the article \cite{crawford} has come to
the authors' attention. This paper studies the dependence of various coherent
states on the weighting parameters $\{\rho_n\}$ and how they effect various
correlation functions of interest regarding general systems, and
particularly for the hydrogen atom. The studies reported in \cite{crawford}
offer a good complement to those of the present paper.

\section*{ACKNOWLEDGEMENTS}

JPA and JRK are pleased to acknowledge the hospitality of the Laboratoire de Physique Th\'eorique de la Mati\`ere
Condens\'ee, Universit\'e Paris 7 -- Denis Diderot during a part of the preparation of this work.  As for JPG, he
acknowledges the hospitality of the Institut de Physique Th\'eorique, Universit\'e Catholique de Louvain,
 Louvain-la-Neuve. KAP thanks J. M. Sixdeniers for his efficient collaboration. 
 Finally, we all thank Achim Kempf for his constructive comments and suggestions.


\newpage

\section*{Figure Captions}

\underline{Figure 1} : 
The infinite square-well potential.\\

\noindent\underline{Figure 2} : 
The  P\"{o}schl--Teller potential 
$V(x) = \frac{1}{2} V_o \left[ {\lambda (\lambda-1)}{\cos^{-2} \frac{x}{2a}} + 
  {\kappa (\kappa -1)}{\sin^{-2}\frac{x}{2a}} \right],$ with $a =  \pi^{-1}$  
 and for $(\lambda,\kappa) = (4,4), (4,8), (4,16)$ (from  bottom to top). \\ 

\noindent\underline{Figure 3} : 
The position $x(t)$ of the particle trapped in the infinite square-well of width $\pi a$, as a function of time. \\ 

\noindent\underline{Figure 4} : The velocity $\mbox{\boldmath $v$} (t)$ of the particle in the infinite square-well:
periodized Haar function.
\\

\noindent\underline{Figure 5} : The acceleration \mbox{\boldmath $\gamma$(t)} of the particle
of the particle in the infinite square-well.
\\

\noindent\underline{Figure 6} : Phase trajectory of the particle in the infinite square-well. \\

\noindent\underline{Figure 7} : The position $x(t)$ of the particle in the symmetric P\"{o}schl--Teller potential
$\lambda =\kappa = 2$: (a) $E=8 V_o, T = \frac{\pi}{2}$;  and (b) $E=16 V_o, T = \frac{\pi}{2\sqrt{2}}$
(compare   Figure \ref{figure3}).
\\

\noindent\underline{Figure 8} : The velocity $\mbox{\boldmath $v$} (t)$ of the particle in the symmetric (2,2) 
P\"{o}schl--Teller potential,
 for the same values of $E$ and $T$ as in  Figure \ref{figure7}
(compare  Figure \ref{figure4}).
\\

\noindent\underline{Figure 9} : The acceleration \mbox{\boldmath $\gamma$}$(t)$ of the particle in the symmetric (2,2)
P\"{o}schl--Teller potential, for the same values of $E$ and $T$ as in  Figure \ref{figure7}
(compare   Figure \ref{figure5}).
\\

\noindent\underline{Figure 10} : Upper part of the phase trajectory of the particle  
the symmetric (2,2) P\"{o}schl--Teller system,
 for the same values of $E$ and $T$ as in  Figure \ref{figure7}
(compare   Figure \ref{figure6}).
\\

\noindent\underline{Figure 11} : (a) The weighting distribution  $|c_n |^2 \equiv  D(n,J, \nu )   $ 
given in (\ref{distrPT}) for 
the infinite square-well  $\nu = 2$ and different
values of $J$. Note the almost Gaussian shape at $J=300$,  centered at 
$n = \langle n \rangle =\sqrt{J} -\frac\nu 2 - \frac 14 \simeq 16$, a  width equal to 
$2 \Delta n =   \sqrt{2}  J^{1/4}\simeq  5.9$;
 (b) The same for for the harmonic oscillator: $|c_n |^2 = \frac {1}{n!} |\alpha|^{2n} e^{-|\alpha|^{2}}.$ 
The values of $\alpha$ are chosen so 
as to get essentially the same mean energy values as in (a): $\alpha = \sqrt{J}$.
\\

\noindent\underline{Figure 12} : The evolution (vs. $\gamma$)
of the probability density $|\langle x | J,\gamma \rangle|^2$, in the case of the infinite square-well
for (a) $J= 2$; (b) $J= 10$; and (c) $J= 50$. 
We note the perfect revival at $\gamma = \pi = \frac 12 t_{\rm rev}$ (in suitable units), 
symmetrically  with respect to the center of the well. 
\\

\noindent\underline{Figure 13} :  Squared modulus $|\langle J,0\, |\, J, \omega t  \rangle |^2$ of the
autocorrelation vs. $\gamma = \omega t$ for the infinite square-well, for $J = 2,10,50$.
As in  Figure \ref{figure12},  the large $J$ regime is characterized by the  occurrence of  fractional revivals.
\\

\noindent\underline{Figure 14} :  Squared modulus $|\langle J,0\, |\, J, \omega t  \rangle |^2$ of the
autocorrelation   for the P\"{o}schl--Teller potential with  $n_{\rm max=10}$, for (a)  $J = 20$; (b) $J =  40$..
\\

\noindent\underline{Figure 15} : Temporal behavior of the average position of the particle in
the infinite square-well (in the Heisenberg picture), 
$\langle J,0\,|\,Q(t)\,|\,J,0\rangle = \langle J, \omega t = \gamma \,|\,Q\,|\,J, \omega t= \gamma \rangle$,
as a function of $\gamma = \omega t $, for $J = 2,10,50$.
\\

\noindent\underline{Figure 16} : Temporal behavior of the average position for the  asymmetric P\"{o}schl--Teller
potential $(\lambda,\kappa) =   (4,8) $ with  $n_{\rm max=10}$, for (a) $J=20$; and (b) $J =  50$.
\\

\noindent\underline{Figure 17} : Temporal behavior of the average momentum $\langle J,0\,| \,P(t) \,|\, J,0\rangle$ in the
case of the infinite square-well, for   $J = 2, 10, 50$. \\ 

\noindent\underline{Figure 18} : Temporal behavior of the squared uncertainty in position $(\Delta Q)^2$,
in the case of the infinite square-well, for   $J = 2, 10, 50$.
\\

\noindent\underline{Figure 19} : Temporal behavior of the squared uncertainty in momentum $(\Delta P)^2$,
in the case of the infinite square-well, for   $J = 2, 10, 50$.
\\

\noindent\underline{Figure 20} : Temporal behavior of the product of the squared uncertainties $(\Delta Q)^2\,(\Delta P)^2$, 
in the case of the infinite square-well, for  $J = 2, 10, 50$.
\\

\noindent\underline{Figure 21} : Temporal behavior of the average position in the case of the infinite square-well, 
for a very high value    $J =10^6$.
 
\end{document}